\documentclass[11pt,a4paper]{article}
\pdfoutput=1

\usepackage[inline]{enumitem}
\usepackage{etex}
\usepackage{comment}
\usepackage{mathrsfs}
\usepackage{multirow}
\usepackage{booktabs}
\usepackage{tabularx}
\usepackage{array}
\usepackage{slashed}
\usepackage{float}
\usepackage{leftidx}
\usepackage{setspace}
\usepackage{verbatim}
\usepackage{adjustbox}
\usepackage{tikz}
\usepackage[caption=false]{subfig}
\usepackage{arydshln}
\usepackage{jheppub}
\usepackage{geometry}
\usepackage{amsthm}

\numberwithin{equation}{section}
\usetikzlibrary{arrows,decorations.markings,shapes.arrows,patterns,positioning}

\newcommand{\bea}{\begin{eqnarray}}
\newcommand{\eea}{\end{eqnarray}}
\newcommand{\bean}{\begin{eqnarray*}}
\newcommand{\eean}{\end{eqnarray*}}

\newcommand{\blue}{\color{blue}}

\def\W #1{\widetilde{#1}}

\def\Label#1{\label{#1}}
\renewcommand{\eqref}[1]{eq.~(\ref{#1})}

\newcolumntype{C}{>{\centering\arraybackslash}X}

\def\braket#1{\left\langle #1 \right\rangle}

\newcommand{\ctobedelete}[1]{}

\DeclareMathOperator{\rank}{rank}

\theoremstyle{definition}

\newtheorem{thm}{Theorem}[section]

\hypersetup{pageanchor=false}
\title{Characterizing the solutions to scattering equations that support tree-level $\text{N}^{k}\text{MHV}$ gauge/gravity amplitudes}
\author[a]{Yi-Jian Du,}
\emailAdd{yijian.du@whu.edu.cn}
\author[c]{Fei Teng}
\emailAdd{Fei.Teng@utah.edu}
\author[b,c]{and Yong-Shi Wu}
\emailAdd{wu@physics.utah.edu}
\affiliation[a]{Center for Theoretical Physics,
School of Physics and Technology,
Wuhan University,\\
299 Bayi Road, Wuhan 430072,
China}
\affiliation[b]{Department of Physics and
Center for Field Theory and Particle Physics, Fudan University,\\
220 Handan Road, Shanghai 200433, China}
\affiliation[c]{Department of Physics and Astronomy, University of Utah,\\ 115 South 1400 East, Salt Lake City, UT 84112, USA}

\abstract{In this paper we define, {independent of theories}, two discriminant matrices involving a solution to the scattering equations in four dimensions, the ranks of which are used to divide the solution set into a disjoint union of subsets.  We further demonstrate, {entirely within the Cachazo-He-Yuan formalism,} that each subset of solutions gives nonzero contribution to tree-level $\text{N}^{k}\text{MHV}$ gauge/gravity amplitudes only for a specific value of $k$. Thus the solutions can be characterized by the rank of their discriminant matrices, which in turn determines the value of $k$ of the $\text{N}^{k} \text{MHV}$ amplitudes a solution can support. As another application of the technique developed, we show analytically that in Einstein-Yang-Mills theory, if all gluons have the same helicity, the tree-level single-trace amplitudes must vanish. 
}

\begin{document}
\maketitle
\hypersetup{pageanchor=true}

\section{Introduction}
The Cachazo-He-Yuan (CHY) formalism for scattering amplitudes~\cite{Cachazo:2013gna,Cachazo:2013hca,Cachazo:2013iea}, since its birth, has generated tremendous interests in the amplitude community. It associates the kinematics of massless particles to the punctures of a Riemann sphere, the positions of which are determined by the scattering equations. A massless quantum field theory is defined by specifying a CHY integrand on the punctured Riemann sphere, and the on-shell amplitudes can be obtained through a contour integral, resulting in a summation over the values of the CHY integrand at the punctures determined by the solutions to scattering equations. This new formulation has succeeded in reformulating, at tree-level, Yang-Mills, pure gravity, Einstein-Yang-Mills, $\phi^{3}$ scalar theory, nonlinear sigma model, Dirac-Born-Infeld theory, to name a few~\cite{Cachazo:2014nsa,Cachazo:2014xea}. The extension to loop level has also been intensively studied~\cite{Geyer:2015bja,Adamo:2013tsa,Casali:2014hfa,Adamo:2015hoa,Ohmori:2015sha,Baadsgaard:2015hia,He:2015yua,Geyer:2015jch,Cachazo:2015aol,Feng:2016nrf,Cardona:2016wcr,Geyer:2016wjx}.

The significance of the CHY formalism lies in that it enables direct evaluation of generic $n$-point amplitudes with arbitrary helicity configurations, with no reference to any Feynman diagrams, while the results are equivalent to the summation over an incredibly huge number of Feynman diagrams at large $n$. Practically, it sheds light on revealing a compact formula for generic amplitudes, hidden behind Feynman diagrams and recursive evaluations. It also strongly signals a possible new formulation of quantum field theories over the (punctured) Riemann {surface}, which makes manifest the hidden simplicity in the field theories. 

Despite the progresses and prospects, there are challenges in understanding this formalism. One immediate question is how to solve the scattering equations. Up to now, out of the $(n-3)!$ solutions~\cite{Cachazo:2013gna,Dolan:2015iln,Cardona:2015ouc} there are only two special ones bearing an analytic expression~\cite{Weinzierl:2014vwa,Roberts:1972,FairlieRoberts:1972,Fairlie:2008dg,Monteiro:2013rya}, while all the others are rather complicated and can only be obtained numerically. Based on the two special solutions, the present authors have evaluated directly the $n$-point MHV and anti-MHV ($\overline{\text{MHV}}$) amplitudes in four dimensions for Yang-Mills, gravity and Einstein-Yang-Mills~\cite{Du:2016blz,Du:2016wkt}, putting them naturally into simple and compact forms. For general amplitudes, it has been shown that after a reduction procedure, the summation of the solutions can be associated to coefficients of some multivariate polynomials, so that the analytic forms of the solutions are not needed~\cite{Cachazo:2015nwa,Baadsgaard:2015ifa,Sogaard:2015dba,Lam:2016tlk,Huang:2016zzb,Bjerrum-Bohr:2016juj,Bosma:2016ttj,Zlotnikov:2016wtk,Cardona:2016gon}. 

A remarkable but somewhat intruiging feature of the CHY formalism is that though the scattering equations have only external momenta as input, their solutions seem to contain some amount of dynamical information as well. By now it is a folklore in the community that the solutions to the scattering equations consist of disjoint subsets, each of which contributes to only some specific sector of the gauge and gravity amplitudes. Inspired by the work of Witten~\cite{Witten:2003nn}, Roiban, Spradlin and Volovich~\cite{Roiban:2004yf} in 4d twistor string, Cachazo, He and Yuan~\cite{Cachazo:2013iaa} first revealed that the scattering equations fall into disjoint sectors, whose number of solutions demonstrates an Eulerian number pattern. In addition, the solution sector should correspond to $\text{N}^{k}\text{MHV}$ amplitudes in Yang-Mills theory and gravity. This correspondence between solutions to scattering equations and $\text{N}^{k}\text{MHV}$ amplitudes was also encoded in the ambitwistor approach for gauge and gravity amplitudes~\cite{Geyer:2014fka}. The Eulerian number pattern is further demonstarted in a more recent work \cite{Cachazo:2016sdc} for scalar blocks. 
Review of solutions in 4d and their relationship with $\text{N}^{k}\text{MHV}$ amplitudes can also be found in \cite{He:2016vfi}.

On the other hand, as already shown in \cite{Du:2016blz,Du:2016wkt}, it is worthwhile to develop direct evaluations of scattering amplitudes entirely within the CHY formalism. Along this direction, when trying to study amplitudes beyond MHV from the (integrated) CHY formula, one encounters again the correspondence between the solutions to scattering equations and the helicity configurations for amplitudes. This calls for a straightforward  characterization of solutions to scattering equations and for the study of their relation to the CHY integrand in $\text{N}^{k}\text{MHV}$ configuration, without using any twistor or ambitwistor argument. The advantage of doing this is that if succeeds, there is a chance that we gain new insights into quantum field theories constructed in arbitrary dimensions, as the discovery of scattering equations has taught us so far.

In this work, we will start from the CHY integrand directly and use elementary transformations of matrices, to answer why and how the solutions know at which $k$ they would support the tree-level $\text{N}^{k}\text{MHV}$ gauge and gravity amplitudes in four dimensions. The total helicity information is packed into a Pfaffian in the CHY integrand, so that we need to understand why this Pfaffian is evaluated to zero. We find that at a given $k$, the Pfaffian is nonzero only if one of its submatrices has a specific rank. Motivated by this observation, we define two discriminant matrices which are independent of theories. Using the rank of either matrix, we can give the solution set a partition such that each subset can only support the amplitudes with one specific $k$. In addition, there is a one-to-one correspondence between $k$ and the rank.
We are going to show more details on this intuition in Sec.~\ref{sec:intuition}. This rank characterization classifies solutions by a set of algebraic equations consistent with the scattering equations. It will be interesting to further study why the number of solutions that satisfy both our new sets of algebraic equations and the scattering equations equals exactly to an Eulerian number. 


Using the above techniques we developed, we further find that in tree-level single-trace Einstein-Yang-Mills amplitudes, if the graviton helicity configuration is fixed, we know the range of $k$ for which the $\text{N}^{k}\text{MHV}$ amplitudes are nonzero. In particular, we prove that the tree-level single-trace Einstein-Yang-Mills amplitudes with gluons having the same helicity must vanish identically. This statement first appeared as a conjecture in~\cite{Bern:1999bx}. We can promote it into a theorem, which suggests that the CHY formalism has indeed captured some fundamental features of field theory that have been obscured in the Lagrangian formalism. 

We note that our derivation is carried out in four dimensions using the spinor helicity formalism~\cite{Xu:1986xb}. However, there is numerical evidence showing that the previous-mentioned $\text{N}^{k}\text{MHV}$ characterization applies in arbitrary dimensions. Thus we conjecture that by proper generalization, our rank characterization could also work in other dimensions.

The layout of this paper is as follows. In Sec.~\ref{sec:SE} we review the scattering equations and the CHY integrands involved in various theories. In Sec.~\ref{sec:intuition}, we  motivate and sketch our approach that starts from newly defined discriminant matrices. The main theorem and other details of our rank characterization of the solutions to scattering equations are presented in Sec.~\ref{sec:chara}. Then we prove in Sec.~\ref{sec:solhel} that {the knowledge of the rank of the discriminant matrices is strong enough to make} the CHY integrand nonzero only at one helicity configuration, so that {the rank associated with a solution naturally dictates the helicity configuration of the amplitude that the solution supports}. In Sec.~\ref{sec:generalamp}, we apply techniques to the Einstein-Yang-Mills theory and show that the interplay between the solutions and helicity configurations lead to the vanishing of gluon-same-helicity amplitudes. {Sec.~\ref{sec:numerical} is devoted to} numerical {verification of our main theorem with examples}, which in addition demonstrate the Eulerian number pattern for the numbers of solutions in the subsets labled by their discriminant rank. Finally, after presenting in Sec.~\ref{sec:dis} an argument that our discriminant matrices can be viewed as gauge equivalent to the matrices given in the twistor-string approach in refs. \cite{Cachazo:2012pz}, we conclude our study {with discussions} in Sec.~\ref{sec:con}. {Some details of our calculation are put} into Appendix~\ref{sec:proof}. 

\section{Scattering equations and CHY formalism}\label{sec:SE}
The backbone of the CHY formalism is the scattering equations for $n$ massless particles, labeled by $a$:
\begin{align}
&\sum_{\substack{b=1\\b\neq a}}^{n}\frac{s_{ab}}{z_{ab}}=0\,,& a\in\mathsf{p}=\{1,2,\ldots,n\}\,,
\Label{eq:SE}
\end{align}
where $s_{ab}=2k_{a}\cdot k_{b}$ are the Mandelstam variables and $z_{ab}\equiv z_{a}-z_{b}$. The number of solutions to \eqref{eq:SE} is $(n-3)!$~\cite{Cachazo:2013gna,Dolan:2015iln,Cardona:2015ouc}. In the following, we use $\{\sigma\}$ to denote one general solution. The $n$-point massless tree-level amplitudes are supported by the solutions of \eqref{eq:SE}, which can be written formally as:
\begin{equation}
	A_{n}=\sum_{\{\sigma\}\in\text{sol.}}\frac{\mathcal{I}_{n}(\sigma)}{{\det}'[\pmb{\Phi}(\sigma)]}\,.
\Label{eq:CHYsum}
\end{equation}
The CHY integrand $\mathcal{I}_{n}$, once elegantly chosen, can reproduce tree-level amplitudes of various theories. The denominator $\det'(\pmb{\Phi})$ is the Jacobian resulting from a contour integral on the $z_{a}$ space, whose specific form is not relevant to this work. We refer the readers to the original CHY paper~\cite{Cachazo:2013hca} for its expression. 

In a variety of theories, the CHY integrand $\mathcal{I}_{n}$ contains the reduced Pfaffian $\text{Pf}\,'(\Psi)$. As a non-inclusive list, we have:\footnote{Comparing with the original definition in~\cite{Cachazo:2013hca}, we have extracted a factor $(-\sqrt{2})^{n}$ from the reduced Pfaffian.}
\begin{align}
&\text{Yang-Mills~\cite{Cachazo:2013hca}:}& &\mathcal{I}_{n}=\frac{(-\sqrt{2})^{n}\text{Pf}\,'(\Psi)}{\sigma_{12}\sigma_{23}\ldots\sigma_{n1}}\,,\nonumber\\
&\text{Pure Gravity~\cite{Cachazo:2013hca}:}& &\mathcal{I}_{n}=2^{n}\left[\text{Pf}\,'(\Psi)\times\text{Pf}\,'(\Psi)\right]\,,\nonumber\\
&\text{Singel-trace Einstein-Yang-Mills~\cite{Cachazo:2014nsa}:}& &\mathcal{I}_{n}(s,r)=\frac{(-\sqrt{2})^{n+s}\text{Pf}\,(\Psi_{\mathsf{h}})\text{Pf}\,'(\Psi)}{\sigma_{g_{1}g_{2}}\sigma_{g{2}g_{3}}\ldots\sigma_{g_{r}g_{1}}}\,.
\Label{eq:theories}
\end{align}
In the Einstein-Yang-Mills theory, the integrand is for $s$ external gravitons and $r$ gluons. In the denominator, $g_{i}$ stands for a gluon index. We further note that we only study four dimensional theories in this work, embracing the full power of the spinor helicity formalism.

The quantity $\text{Pf}\,'(\Psi)$ is of particular interests. 
It is defined as:
\begin{align}
	&\text{Pf}\,'(\Psi)=\frac{(-1)^{i+j}}{\sigma_{ij}}\text{Pf}\,(\Psi_{ij}^{ij})\,,& &(1\leqslant i<j\leqslant n)\,,
\Label{eq:rf}
\end{align}
where $\Psi_{ij}^{ij}$ denotes the submatrix of $\Psi$ with the $i$-th and $j$-th row and column deleted. For definiteness, we choose $i=n-1$ and $j=n$ in all the following calculations, and define for convenience:
\begin{align}
&\pmb{\psi}\equiv{\Psi}_{n-1,n}^{n-1,n}\,,& &\text{Pf}\,'(\Psi)=\frac{-1}{\sigma_{n-1,n}}\,\text{Pf}\,(\pmb{\psi})\,.
\end{align}
In the following, we choose the polarization vectors as:
\begin{align}
&\epsilon_{a}^{\mu}(-)=\frac{\langle a|\gamma^{\mu}|q]}{\sqrt{2}[qa]}\,,& &\epsilon_{a}^{\mu}(+)=\frac{\langle p|\gamma^{\mu}|a]}{\sqrt{2}\langle pa\rangle}\,,
\Label{eq:gaugechoice}
\end{align}
which applies to all the particles involved in the scattering amplitudes. In Yang-Mills, $\epsilon_{a}$ describes the polarization of an external gluon. In pure gravity, the polarization of a graviton is given by $\epsilon_{a}^{\mu\nu}=\epsilon_{a}^{\mu}\epsilon_{a}^{\nu}$. For split $\text{N}^{k}\text{MHV}$ amplitudes of both Yang-Mills and gravity:
\begin{equation}
(\,\underbrace{--\ldots-}_{k+2}\,\underbrace{++\ldots+}_{n-k-2}\,)\,,
\label{eq:split}
\end{equation}
the matrix $\Psi$ is given by:
\begin{equation}
  {\Psi}=\left(\begin{array}{cc}
    A & -C^{T} \\
    C & B \\
    \end{array}\right)=\left(\begin{array}{ccc}
{A} & -{C}_{-}^{T} & -{C}_{+}^{T} \\
{C}_{-} & 0 & \mathcal{B} \\
{C}_{+} & -\mathcal{B}^{T} & 0 \\
\end{array}\right)\,,
\Label{eq:Psi}
\end{equation}
in which the blocks have the following forms:
\begin{itemize}
	\item the $n\times n$ matrix ${A}$ is given by:
	\begin{align}
	&A_{ab}=\frac{s_{ab}}{\sigma_{ab}}=-\frac{\braket{ab}[ab]}{\sigma_{ab}}\,,& &(a,b\in\mathsf{p}\text{ and }a\neq b)\,.\Label{eq:A}
	\end{align}
        The diagonal elements of $A$ are defined to be zero: $A_{aa}=0$.
        \item the $n\times n$ matrix $B$ is given by:
        \begin{align}
        &B_{ab}=\frac{\epsilon_{a}\cdot\epsilon_{b}}{\sigma_{ab}}\,,& &(a,b\in\mathsf{p}\text{ and }a\neq b)\,,
        \end{align}
        and the diagonal elements are zero: $B_{aa}=0$. In the gauge~(\ref{eq:gaugechoice}), the antisymmetric matrix $B$ has  a block form, with two diagonal zero blocks:
        \begin{equation*}
          B=\left(\begin{array}{cc}
            0 & \mathcal{B} \\
            -\mathcal{B}^{T} & 0 \\
            \end{array}\right)\,,
        \end{equation*}
        where the $(k+2)\times(n-k-2)$ matrix $\mathcal{B}$ is given by:
	\begin{align}
	&\mathcal{B}_{ab}=\frac{\epsilon_{a}(-)\cdot\epsilon_{b}(+)}{\sigma_{ab}}=\frac{\braket{ap}[bq]}{[aq]\braket{bp}\sigma_{ab}}\,,& &(a\in\mathsf{p}^{-},\;b\in\mathsf{p}^{+})\,.\Label{eq:B}
	\end{align}
	We use $\mathsf{p}^{-}$ $(\mathsf{p}^{+})$ to stand for the set of negative (positive) helicity particles, with size:
	\begin{equation*}
	|\mathsf{p}^{-}|=k+2\,,\qquad |\mathsf{p}^{+}|=n-k-2\,,\qquad (0\leqslant k\leqslant n-2).
	\end{equation*}
        \item the $n\times n$ matrix $C$ is given by:
        \begin{equation}
	C_{ab}=\left\{\begin{array}{>{\displaystyle}l @{\hspace{2em}\vspace{0.5em}} >{\displaystyle}l}
	-\frac{\sqrt{2}\epsilon_{a}\cdot k_{b}}{\sigma_{ab}} & a\neq b\\ 
	\sum_{c\neq a}\frac{\sqrt{2}\epsilon_{a}\cdot k_{c}}{\sigma_{ac}} & a=b \\ 
	\end{array}\right.\,.
        \Label{eq:C}
        \end{equation}
        In the gauge (\ref{eq:gaugechoice}), $C$ can be separated into two parts:
        \begin{equation*}
          C=\begin{pmatrix}
          C_{-} \\ C_{+} \\
          \end{pmatrix}\,,
        \end{equation*}
        where the $(k+2)\times n$ matrix $C_{-}$ is given by:
	\begin{equation}
	(C_{-})_{ab}=\frac{\langle ab\rangle}{\sigma_{ab}}\frac{[bq]}{[aq]}\,,\qquad a\in\mathsf{p}^{-}\,;
	\Label{eq:C-}
	\end{equation}
	and the $(n-k-2)\times n$ matrix $C_{+}$ is given by:
	\begin{equation}
	(C_{+})_{ab}=\frac{[ab]}{\sigma_{ab}}\frac{\braket{bp}}{\braket{ap}}\,,\qquad a\in\mathsf{p}^{+}\,.
	\Label{eq:C+}
	\end{equation}
\end{itemize}
We further assume that $p$ and $q$ do not coincide with any external momenta such that the matrices $A$, $B$ and $C$ do not contain additional zero elements except for those mentioned before.

For split $\text{N}^{k}\text{MHV}$ Einstein-Yang-Mills amplitudes, $\Psi$ has the same definition as (\ref{eq:Psi}), which contains the gluon polarization $\epsilon_{g_a}$ and half of the graviton polarization, say, $\epsilon_{h_a}^{\mu}$ of $\epsilon_{h_a}^{\mu\nu}=\epsilon_{h_a}^{\mu}\epsilon_{h_a}^{\nu}$. The other half of the graviton polarization, say $\epsilon_{h_a}^{\nu}$, is contained in the $2s\times 2s$ matrix $\Psi_{\mathsf{h}}$, having the same structure as $\Psi$:
\begin{equation}
{\Psi}_{\mathsf{h}}=\left(\begin{array}{ccc}
{A}_{\mathsf{h}} & -({C}_{\mathsf{h}})_{-}^{T} & -({C}_{\mathsf{h}})_{+}^{T} \\
({C}_{\mathsf{h}})_{-} & 0 & \mathcal{B}_{\mathsf{h}} \\
({C}_{\mathsf{h}})_{+} & -\mathcal{B}_{\mathsf{h}}^{T} & 0 \\
\end{array}\right)\,.
\Label{eq:Psih}
\end{equation}
The components of $\Psi_{\mathsf{h}}$ are:
\begin{itemize}
	\item the $s\times s$ matrix $A_{\mathsf{h}}$ has the same definition as (\ref{eq:A}), but with index $a$ and $b$ ranging within the graviton set $\mathsf{h}$.
	\item the $s^{-}\times s^{+}$ matrix $\mathcal{B}$ has the same definition as (\ref{eq:B}), but with the row index $a$ ranging within the negative helicity graviton set $\mathsf{h}_{-}$ (whose order is $s^{-}$) and the column index $b$ within the positive graviton set $\mathsf{h}_{+}$ (whose order is $s^{+}$).
	\item the $s^{+}\times s$ matrix $(C_{\mathsf{h}})_{+}$ and the $s^{-}\times s$ matrix $(C_{\mathsf{h}})_{-}$ have the same definitions as (\ref{eq:C-}) and (\ref{eq:C+}), but with column indices ranging within the graviton set $\mathsf{h}$ and row indices within $\mathsf{h}_{\pm}$ respectively.
\end{itemize}

Since the information of total helicity configuration is encoded in $\text{Pf}\,'(\Psi)$ in all those theories in (\ref{eq:theories}), its support must agree with that of $\text{N}^{k}\text{MHV}$ amplitudes. 
In addition, $\text{Pf}\,'(\Psi)$ is invariant under simultaneous permutations of rows and columns such that at $\text{N}^{k}\text{MHV}$, it is sufficient to study only the split helicity configuration (\ref{eq:split}).

\section{Relating rank to helicity configuration: intuition and general discussion}\label{sec:intuition}
For the $\text{N}^{k}\text{MHV}$ amplitudes, a natural question is whether all solutions to the scattering  equations support (or contribute to) them or not. Numerical calculations for small $n$ provide evidence that only a subset of solutions to scattering equations~(\ref{eq:SE}) should be included in the summation of \eqref{eq:CHYsum} for the $\text{N}^{k}\text{MHV}$ amplitudes, because the contribution from the others actually vanish. We will prove below analytically that indeed this is generally true. For more detailed understanding, one may ask: Could we give a characterization of the subset of solutions that support the  $\text{N}^{k}\text{MHV}$ amplitudes, without knowing explicitly the solutions? Do the subsets of solutions supporting the $\text{N}^{k}\text{MHV}$ and the $\text{N}^{k'}\text{MHV}$ amplitudes with $k\neq k'$, respectively, overlap or not? In this paper we will present our analytic study and our answer to these questions. 

The following two $n\times n$ discriminant matrices are of paramount importance in characterizing the solutions:
\begin{align}
&(\mathfrak{C}_{-})_{ab}=\left\{\begin{array}{>{\displaystyle}l @{\hspace{1em}} >{\displaystyle}l}
\frac{\langle ab\rangle}{\sigma_{ab}} & a\neq b\\
-\Sigma_{a}^{-} & a=b \\
\end{array}\right.\,,& &\Sigma_{a}^{-}=\sum_{\substack{b=1 \\ b\neq a}}^{n}\frac{\langle ab\rangle[bq]}{\sigma_{ab}[aq]}\,,& &(a,b\in\mathsf{p})\,,\nonumber\\
&(\mathfrak{C}_{+})_{ab}=\left\{\begin{array}{>{\displaystyle}l @{\hspace{1em}} >{\displaystyle}l}
\frac{[ab]}{\sigma_{ab}} & a\neq b\\
-\Sigma_{a}^{+} & a=b \\
\end{array}\right.\,,& &\Sigma_{a}^{+}=\sum_{\substack{b=1 \\ b\neq a}}^{n}\frac{[ab]\langle bp\rangle}{\sigma_{ab}\langle ap\rangle }\,,& &(a,b\in\mathsf{p})\,.
\Label{eq:Cpm}
\end{align}
We note that the diagonal elements actually do not depend on the reference spinors $|p\rangle$ and $|q]$. The reason is as follows. If we have another spinor $|\tilde{q}]$, we can show that:
\begin{align}
\sum_{b\neq a}\frac{\langle ab\rangle[b\tilde{q}]}{\sigma_{ab}[a\tilde{q}]}=\sum_{b\neq a}\frac{\langle ab\rangle[b\tilde{q}][aq]}{\sigma_{ab}[a\tilde{q}][aq]}=\sum_{b\neq a}\left(\frac{\langle ab\rangle[ab][q\tilde{q}]}{\sigma_{ab}[a\tilde{q}][aq]}+\frac{\langle ab\rangle[bq]}{\sigma_{ab}[aq]}\right)=\sum_{b\neq a}\frac{\langle ab\rangle[bq]}{\sigma_{ab}[aq]}=\Sigma_{a}^{-}\,,
\end{align}
where we have used the scattering equation to obtain the third equality. Similar identity also holds for $\Sigma_{a}^{+}$. One of the main results of our work is that $\rank(\mathfrak{C}_{\pm})$ decides whether $\text{Pf}\,'(\Psi)$ vanishes at $\text{N}^{k}\text{MHV}$ helicity configurations. Meanwhile, different solutions may give $\mathfrak{C}_{\pm}$ different ranks. Therefore, the importance of $\rank(\mathfrak{C}_{\pm})$ is twofold:
\begin{itemize}
	\item The rank condition serves as an algebraic classification of the solutions, which divides the solution set into disjoint unions.
	\item The quantity $\rank(\mathfrak{C}_{\pm})$ serves as the bridge connecting $\text{N}^{k}\text{MHV}$ helicity configurations to solutions. It is the key to understand why one certain solution gives nonzero contribution to amplitudes only at one $k$. 
\end{itemize}
Before delving into detailed calculation, we would like to give readers an intuition on why the rank of $\mathfrak{C}_{\pm}$, containing no information of the helicities, can dictate whether $\text{Pf}\,'(\Psi)$ vanishes or not at a certain helicity configuration.


To start with, it is important to realize that helicity configuration decides the dimension of the matrix $\mathcal{B}$ and how we glue $\mathfrak{C}_{\pm}$, together with proper gauge dependence, into the $C$ matrix: 
\begin{align}
&(C_{-})_{ab}=(\mathfrak{C}_{-})_{ab}\frac{[bq]}{[aq]}\,,\quad a\in\mathsf{p}^{-}\,,\nonumber\\ &(C_{+})_{ab}=(\mathfrak{C}_{+})_{ab}\frac{\langle bp\rangle}{\langle ap\rangle}\,,\quad a\in\mathsf{p}^{+}.
\Label{eq:C-relation}
\end{align}
The rank structure of $C$, namely, whether $C_{\pm}$ part has rank deficiency or not, is determined by both the helicity configuration and $\rank(\mathfrak{C}_{\pm})$. 
{When calculating minors of $C_{\pm}$, the reference spinor part can be pulled out of the determinant, since $[aq]$ and $[bq]$ are nonzero common factors in $a$-th row and $b$-th column. Schematically, we have:
\begin{align*}
&\text{minor}\,(C_{-})=\left(\prod_{\text{row}}\frac{1}{[aq]}\right)\left(\prod_{\text{column}}[bq]\right)\text{minor}\,(\mathfrak{C}_{-})\,,\\
&\text{minor}\,(C_{+})=\left(\prod_{\text{row}}\frac{1}{\langle ap\rangle}\right)\left(\prod_{\text{column}}\langle bp\rangle\right)\text{minor}\,(\mathfrak{C}_{+})\,.
\end{align*}
As a result, when a minor of $\mathfrak{C}_{\pm}$ vanishes, so does the corresponding one of $C_{\pm}$, which leads to the fact that if $\rank({C}_{\pm})$ is smaller than the row number, it must equal to $\rank({\mathfrak{C}}_{\pm})$. Consequently, we have:
\begin{align}
\label{eq:rankrel}
&\text{rank}(C_{-})=\min\left\{\,k+2,\;\text{rank}(\mathfrak{C}_{-})\,\right\}\,,& &\text{rank}(C_{+})=\min\left\{\,n-k-2,\;\text{rank}(\mathfrak{C}_{+})\,\right\}\,.
\end{align}
The row number $k+2$ and $n-k-2$ also come into play since it is the upper bound of $\rank{C}_{\pm}$.} Suppose $C_{-}$ has rank $r$, we can always find an $r\times r$ submatrix $\mathcal{R}$ with nonzero determinant. Then using proper elementary transformations, we can make zero all elements of $\pmb{\psi}$ that share either rows or columns with $\mathcal{R}$ (and $-\mathcal{R}^{T}$ in the symmetric location). As a result, $\text{Pf}\,'(\Psi)$ can be evaluated by:
\begin{equation}
\text{Pf}\,'(\Psi)\propto\text{Pf}\,(\pmb{\psi})=\pm\det(\mathcal{R})\times\text{Pf}\,(\W{\pmb{\psi}})\,,
\end{equation}
where $\W{\pmb{\psi}}$ is $(2n-2-2r)\times(2n-2-2r)$ dimensional, and the $\pm$ sign is determined by the position of $\mathcal{R}$. Now the matrix $\W{\pmb{\psi}}$ clearly does exhibit an interplay between the rank $r$ and the helicity configuration: the overall size of $\W{\pmb{\psi}}$ depends on $r$ while its structure, inherited from $C_{\pm}$ and $\mathcal{B}$, depends on $k$. If we choose $\mathcal{R}$ in a smart way, $\W{\pmb{\psi}}$ may have a good structure (namely, large zero blocks) for ones to tell that it can be nondegenerate only at certain helicity configurations, depending on $r$.

The above discussion serves as our intuition that guides the following calculation. In Sec.~\ref{sec:chara}, we first show that different solutions to the scattering equations can lead to a spectrum of $\rank(\mathfrak{C}_{\pm})$. Then in Sec.~\ref{sec:solhel}, we demonstrate with detailed calculation that the idea in the previous paragraph can indeed be realized. Therefore, $\rank(\mathfrak{C}_{\pm})$ links the solutions to the helicity configurations.

\section{Characterizing the solutions to the scattering equations by rank}\label{sec:chara}
In this section, we establish the characterization of the solutions to the scattering equations by the rank of $\mathfrak{C}_{\pm}$.
We also propose a theorem that relates the rank to the helicity configuration that this solution can support, which will be proved in Sec.~\ref{sec:solhel}. 
Some numerical results will be given in Sec.~\ref{sec:numerical}.

\subsection{The rank of \texorpdfstring{$\mathfrak{C}_{\pm}$}{C+-} and \texorpdfstring{$C_{\pm}$}{C+-}}
Independent of the solutions, the rank of $\mathfrak{C}_{\pm}$ cannot exceed $n-2$, since they always have two null vectors respectively:
\begin{align*}
	&\left(\,[1q],\,[2q],\,\cdots,\,[nq]\,\right)^{T}\,,\quad\left(\,[1q]\sigma_{1},\,[2q]\sigma_{2},\,\cdots,\,[nq]\sigma_{n}\,\right)^{T}& &\text{for }\mathfrak{C}_{-}\,,\\
	&\left(\,\braket{1p},\,\braket{2p},\,\cdots,\,\braket{np}\,\right)^{T}\,,\quad\left(\,\braket{1p}\sigma_{1},\,\braket{2p}\sigma_{2},\,\cdots,\,\braket{np}\sigma_{n}\,\right)^{T}& &\text{for }\mathfrak{C}_{+}\,.
\end{align*}
In other words, $\mathfrak{C}_{\pm}$ can be of rank $n-2$ at most. Similarly, the rank of $C$ must also be smaller than $n-2$, since it always has two null vectors:
\begin{equation*}
\left(\,1,\,1,\,\cdots,\,1\,\right)^{T}\,,\quad\left(\,\sigma_{1},\,\sigma_{2},\,\cdots,\,\sigma_{n}\,\right)^{T}\,,
\end{equation*}
independent of the solutions~\cite{Cachazo:2013hca}. 

Next, we show that by a proper choice of gauge, we can always make:
\begin{equation}
\rank(C)=\rank(C_{-})+\rank(C_{+})\,.
\Label{eq:rankC}
\end{equation}
If $\rank(C)$ is smaller than the value of (\ref{eq:rankC}), it means that there is at least one linear relation between rows of $C_{-}$ and $C_{+}$. For example, one row of ${C}_{+}$ can be obtained by a superposition of $s$ linearly independent rows of ${C}_{-}$. Should this situation happen, it would be necessary that the following $(s+1)\times(s+1)$ determinant vanishes:
\begin{equation}
\begin{vmatrix}
({C}_{-})_{i_{1}j_{1}} & ({C}_{-})_{i_{1}j_{2}} & \cdots & ({C}_{-})_{i_{1}j_{s+1}} \\
({C}_{-})_{i_{2}j_{1}} & ({C}_{-})_{i_{2}j_{2}} & \cdots & ({C}_{-})_{i_{2}j_{s+1}} \\
\vdots & \vdots & & \vdots \\
({C}_{-})_{i_{s}j_{1}} & ({C}_{-})_{i_{s}j_{2}} & \cdots & ({C}_{-})_{i_{s}j_{s+1}} \\
({C}_{+})_{kj_{1}} & ({C}_{+})_{kj_{2}} & \cdots & ({C}_{+})_{kj_{s+1}} \\
\end{vmatrix}=0\,,
\Label{eq:lineardep}
\end{equation}
while in the ${C}_{-}$ part at least one $s\times s$ minor is nonzero. Then by choosing another reference spinor, say, $|\W{p}\rangle$, in $C_{+}$, we can break the equality of \eqref{eq:lineardep}. Meanwhile, $\text{Pf}\,'(\Psi)$ is gauge invariant so that changing gauge will not modify the value of $\text{Pf}\,'(\Psi)$ for a given solution. Therefore, we can use the gauge freedom to eliminate any potential linear relations as in (\ref{eq:lineardep}) and make \eqref{eq:rankC} hold. Consequently, we have:
\begin{equation}
\label{eq:rank2}
\rank(C)=\rank(C_{-})+\rank(C_{+})\leqslant n-2\,.
\end{equation}

Now we can prove that for all solutions, $\rank(\mathfrak{C}_{\pm})\leqslant n-3$. Should there be a solution that makes $\rank(\mathfrak{C}_{-})=n-2$, we could extract $n-2$ linearly independent rows and make them into $C_{-}$. The helicity configuration is thus anti-MHV while the $C_{+}$ part only contains two rows. However, we must have $\rank(C_{+})\geqslant 1$ such that at anti-MHV the solution that leads to $\rank(\mathfrak{C}_{-})=n-2$ would lead to $\rank(C)\geqslant n-1$, which contradicts with the condition $\rank(C)\leqslant n-2$. The $\rank(\mathfrak{C}_{+})\leqslant n-3$ part of the statement can be proved in the same way with a MHV configuration. 

Among all the $(n-3)!$ solutions, we have two special rational ones with analytical expressions~\cite{Weinzierl:2014vwa,Roberts:1972,FairlieRoberts:1972,Fairlie:2008dg,Monteiro:2013rya}:
\begin{align}
	&\sigma_{a}^{(1)}=\frac{\braket{a,n-2}\braket{n-1,n}}{\braket{an}\braket{n-1,n-2}}& &\sigma_{a}^{(2)}=\frac{[a,n-2][n-1,n]}{[an][n-1,n-2]}\,.
\Label{eq:special}
\end{align}
By plugging them into \eqref{eq:Cpm}, we find that:
\begin{align}
	&\rank[\mathfrak{C}_{-}(\sigma^{(1)})]=1\,,& &\rank[\mathfrak{C}_{-}(\sigma^{(2)})]=n-3\,,\nonumber\\
	&\rank[\mathfrak{C}_{+}(\sigma^{(1)})]=n-3\,,& &\rank[\mathfrak{C}_{+}(\sigma^{(2)})]=1\,.
\end{align}
Thus the minimal rank $1$ and the maximal rank $n-3$ can indeed be achieved by some solution.

\subsection{Characterizing solutions by the rank of \texorpdfstring{$\mathfrak{C}_{\pm}$}{C+-}}
Now we can define a partition of the solution set:
\begin{align}
	&\text{solution set}\,=\bigcup_{m=0}^{n-4}\pmb{P}_{-}(n-3,m)\,,& \pmb{P}_{-}(n-3,i)\cap\pmb{P}_{-}(n-3,j)=\varnothing\;\text{ if }\;i\neq j\,,
\end{align}
such that for all $\{\sigma\}\in\pmb{P}_{-}(n-3,m)$, we have:
\begin{equation}
	\rank[\mathfrak{C}_{-}(\sigma)]=m+1\,.
\Label{eq:classify}
\end{equation}
Based on $\rank(\mathfrak{C}_{+})$, we can derive another partition $\pmb{P}_{+}(n-3,m)$ by a similar construction:
\begin{align}
&\text{solution set}\,=\bigcup_{m=0}^{n-4}\pmb{P}_{+}(n-3,m)\,,& \pmb{P}_{-}(n-3,i)\cap\pmb{P}_{+}(n-3,j)=\varnothing\;\text{ if }\;i\neq j\,,
\end{align}
such that for all $\{\omega\}\in\pmb{P}_{+}(n-3,m)$, we have:
\begin{equation}
\rank[\mathfrak{C}_{+}(\omega)]=m+1\,.
\Label{eq:classify2}
\end{equation}
Since the matrix $\mathfrak{C}_{-}(\sigma)$ and $\mathfrak{C}_{+}(\sigma^{\ast})$ are complex conjugate to each other, they must have the same rank. Therefore, we must have:
\begin{equation*}
	\rank[\mathfrak{C}_{+}(\sigma^{\ast})]=m+1\,,\qquad\{\sigma\}\in\pmb{P}_{-}(n-3,m)\,,
\end{equation*}
which means that $\pmb{P}_{-}^{\ast}(n-3,m)$, the subset composed of all $\{\sigma^{\ast}\}$ with $\{\sigma\}\in\pmb{P}_{-}(n-3,m)$, is a subset of $\pmb{P}_{+}(n-3,m)$. We can derive in the same way that $\pmb{P}_{+}^{\ast}(n-3,m)$ is a subset of $\pmb{P}_{-}(n-3,m)$. The two sets thus must be conjugate to each other:\footnote{
	{We note that this property is the consequence of real Minkowski signature momenta, such that $|i\rangle$ and $|i]$ are conjugate to each other.}}
\begin{equation}
\pmb{P}_{-}^{\ast}(n-3,m)=\pmb{P}_{+}(n-3,m),\qquad \left|\pmb{P}_{-}(n-3,m)\right|=\left|\pmb{P}_{+}(n-3,m)\right|\,.
\Label{eq:Pcc}
\end{equation}
We note that this rank characterization is invariant under the simultaneous rescaling:
\begin{equation}
	|{i}\rangle\,\rightarrow\,t|{i}\rangle,\quad|{i}]\,\rightarrow\,t^{-1}|{i}]\,,
\Label{eq:scale}
\end{equation}
which means it is purely kinematical. The reason is simple: under the rescaling (\ref{eq:scale}), the factor $t$ gets multiplied always to one entire row and column of $\mathfrak{C}_{\pm}$. As a result, when calculating a minor, $t$ can always be pulled out of the determinant so that the rank is unchanged.

For all $\{\sigma\}\in\pmb{P}_{-}(n-3,m)$, we can extract $m+1$ linearly independent rows to make the $C_{-}$ part of the $C$ matrix. 
{Now the $C_{-}$ part has no rank deficiency: $\rank(C_{-})=m+1$. Because of \eqref{eq:rank2}, the rank deficiency of the $C_{+}$ part must be at least two:}
\begin{equation*}
\rank[C_{+}(\sigma)]=\rank[\mathfrak{C}_{+}(\sigma)]\leqslant n-m-3\,.
\end{equation*}
{Here we have $\rank(\mathfrak{C}_{+})=\rank(C_+)$ because $C_+$ must have rank deficiency, according to \eqref{eq:rankrel}.} Therefore, without knowing the specific forms of the solutions, we have shown that the partition $\pmb{P}_{-}$ has the following property: for all $\{\sigma\}\in\pmb{P}_{-}(n-3,m)$, we have:
\begin{align}
	&\rank[\mathfrak{C}_{-}(\sigma)]=m+1\,,& &\rank[\mathfrak{C}_{+}(\sigma)]\leqslant n-m-3\,.
\Label{eq:rankCpm}
\end{align}
Similar result also holds for $\pmb{P}_{+}$. We are now ready to prove a rather strong theorem:
\begin{thm}[\emph{Main Theorem}]
	\label{thm:1}
	Only those solutions in the partition $\pmb{P}_{-}(n-3,k)$ [or $\pmb{P}_{+}(n-3,k)$] can support $\text{Pf}\,'(\Psi)$ at $\text{N}^{k}\text{MHV}$ $(\text{or }\text{N}^{n-k-4}\text{MHV})$ helicity configurations.
\end{thm}
\noindent In other words, the rank partition we have derived implies a characterization labeled by the $\text{N}^{k}\text{MHV}$ amplitudes that a solution can support. 
Theorem~\ref{thm:1} applies to all the theories listed in \eqref{eq:theories}. Actually, this theorem can be applied to all those theories whose CHY integrand is proportional to $\text{Pf}\,'(\Psi)$.

A direct consequence of Theorem~\ref{thm:1} is that $\text{N}^{k}\text{MHV}$ amplitudes are supported by both $\pmb{P}_{-}(n-3,k)$ and $\pmb{P}_{+}(n-3,n-k-4)$, such that they must coincide. Because of \eqref{eq:Pcc}, we can further derive:
\begin{equation}
\pmb{P}_{+}(n-3,n-k-4)=\pmb{P}^{\ast}_{-}(n-3,n-k-4)=\pmb{P}_{-}(n-3,k)\,.
\end{equation}
In other words, the solutions in $\pmb{P}_{-}(n-3,k)$ and $\pmb{P}_{-}(n-3,n-k-4)$ are complex conjugate to each other while these two sets must have the same order. More importantly, the ``$\leqslant$'' of \eqref{eq:rankCpm} is actually ``$=$'':
\begin{align}
&\rank[\mathfrak{C}_{-}(\sigma)]=m+1\,,& &\rank[\mathfrak{C}_{+}(\sigma)]= n-m-3\,,& &\text{for all }\{\sigma\}\in\pmb{P}_{-}(n-3,m)\,.
\Label{eq:rankCpm2}
\end{align}
This equation should be viewed as a corollary of Theorem~\ref{thm:1}, and we emphasize that it plays no role in the proof of Theorem~\ref{thm:1}.
\section{Associating solutions to helicity configurations}\label{sec:solhel}

In this section, we give an analytic proof of Theorem~\ref{thm:1}. Given a helicity configuration, the dimensions of ${C}_{\pm}$ and $\mathcal{B}$ are fixed. However, for different solutions, the rank of ${C}_{\pm}$ may be different. We first study a simple example to demonstrate how the rank affects the value of $\text{Pf}\,'(\Psi)$. 
\subsection{Warm-up: Vanishing of \texorpdfstring{$(+\ldots+)$}{(+...+)} and \texorpdfstring{$(-+\ldots+)$}{(-+...+)} Amplitudes}
Using the gauge (\ref{eq:gaugechoice}), we can easily see that both the helicity configurations $(+\ldots+)$ and $(-+\ldots+)$ lead to $\text{Pf}\,'(\Psi)=0$ such that the amplitudes vanish. For both cases, the matrix $\mathcal{B}$ can be made identically zero: 
\begin{itemize}
	\item For $(+\ldots+)$, an arbitrary $|p\rangle$ will do the job.
	\item For $(-+\ldots+)$, choose $|p\rangle=|i_{-}\rangle$ in \eqref{eq:gaugechoice}, where $|i_{-}\rangle$ is the momentum spinor of the negative helicity particle.
\end{itemize}
Now in $\Psi_{ij}^{ij}$, the lower left $C$ block has dimension $n\times(n-2)$. Since the rows are more than the columns, the rank of $C$ cannot be larger than the column number, which is $n-2$. Therefore, by the definition of the matrix rank, there are at most $n-2$ linear-independent rows in $C$, which we can use to make at least two rows of $C$ zero by elementary transformations. If $C$ has a lower rank than $n-2$, we can make even more rows zero. Since we have $\mathcal{B}=0$ for both cases, now we have two entire rows of $\Psi_{ij}^{ij}$ zero such that its determinant and Pfaffian must be zero. The validity of this operation does not depend on any specific solution. Thus we have proved that $\text{Pf}\,'(\Psi)=0$ for all the solutions.

We present this rather trivial case to give readers a taste that the rank structure of $C$ may control whether $\text{Pf}\,'(\Psi)$ is zero or not.
\subsection{Solutions and \texorpdfstring{$\text{Pf}\,'(\Psi)$}{Pf(Psi)}}\label{sec:pfpsi}
To prove Theorem~\ref{thm:1}, it is sufficient to just show that at $\text{N}^{k}\text{MHV}$, only solutions in $\pmb{P}_{-}(n-3,k)$ can make $\text{Pf}\,'(\Psi)\neq 0$. Eq.~(\ref{eq:rankCpm}) implies that at $\text{N}^{k}\text{MHV}$, those $\{\sigma\}\in\pmb{P}_{-}(n-3,k)$ give $C_{-}$ rank deficiency one and $C_{+}$ more than one. 
In this section, we first focus on those solutions in $\pmb{P}_{-}(n-3,m)$ with $m\leqslant k$, which give $C_{-}$ rank deficiency. 

Theorem~\ref{thm:1} is equivalent to that $\text{Pf}\,(\pmb{\psi})\neq 0$ can only happen for $m=k$. We start with a solution in $\pmb{P}_{-}(n-3,m)$, which gives:
\begin{equation*}
	\text{rank}[{C}_{-}(\sigma)]=m+1\equiv r\,.
\end{equation*}
Then in $C_{-}$, there must be one $r\times r$ submatrix that has a nonzero determinant. Suppose we call this matrix $\mathcal{R}$:
\begin{equation}
	\mathcal{R}=\begin{pmatrix}
		({C}_{-})_{i_1j_1} & ({C}_{-})_{i_1j_2} & \cdots & ({C}_{-})_{i_1j_r} \\
		\vdots & \vdots & & \vdots \\
		({C}_{-})_{i_rj_1} & ({C}_{-})_{i_rj_2} & \cdots & ({C}_{-})_{i_rj_r} 
	\end{pmatrix}\,.
\end{equation}
where rows and columns are chosen from the sets: 
$$\mathsf{i}_{r}=\{i_{1},i_{2},\ldots,i_{r}\}\quad\text{ and }\quad\mathsf{j}_{r}=\{j_{1},j_{2},\ldots,j_{r}\}.$$ Note that $\det(\mathcal{R})$ is proportional to the following $r\times r$ minor of $\mathfrak{C}_{-}$: 
\begin{align}
&\det(\mathcal{R})=\left(\prod_{k=1}^{r}\frac{[j_kq]}{[i_kq]}\right)\det(\mathfrak{C}_{-}^{r})\,,& &\mathfrak{C}_{-}^{r}=\begin{pmatrix}
(\mathfrak{C}_{-})_{i_1j_1} & (\mathfrak{C}_{-})_{i_1j_2} & \cdots & (\mathfrak{C}_{-})_{i_1j_r} \\
\vdots & \vdots & & \vdots \\
(\mathfrak{C}_{-})_{i_rj_1} & (\mathfrak{C}_{-})_{i_rj_2} & \cdots & (\mathfrak{C}_{-})_{i_rj_r} 
\end{pmatrix}\,.
\end{align}
Our strategy is to use $\mathcal{R}$ to make as many elements in ${C}_{-}$ block zero as possible by elementary transformations, found by solving linear relations between rows and columns of ${C}_{-}$.

First, after we permute $\mathcal{R}$ to the upper left corner of $C_{-}$, the shape of $\pmb{\psi}$ before any further manipulation is:
\begin{equation}
\pmb{\psi}=\adjustbox{raise=-2cm}{\begin{tikzpicture}
	\begin{scope}[xshift=-0.0cm,yshift=-0.0cm]
		\path [pattern=north east lines,pattern color=blue!50] (-0.25,-1.25) -- (-2,-1.25) -- (-2,-0.5) -- (-1.25,-0.5) -- (-1.25,0.25) -- (-0.25,0.25) -- cycle;
		\draw [color=black,thick] (-2,-2) rectangle (-0.25,0.25);
		\draw [color=black,thick,dashed] (-2,-0.5) -- (-0.25,-0.5) (-2,-1.25) -- (-0.25,-1.25) (-1.25,-0.5) -- (-1.25,0.25);
		\node at (-1.625,-0.125) [font=\Large]{$\mathcal{R}$};
		\node at (-1.125,-1.625) [font=\Large]{${C}_{+}$};
	\end{scope}
	\begin{scope}[xshift=0.0cm,yshift=0.0cm]
		\path [pattern=north east lines,pattern color=blue!50] (1.25,0.25) -- (1.25,2) -- (0.5,2) -- (0.5,1.25) -- (-0.25,1.25) -- (-0.25,0.25) -- cycle;
		\draw [color=black,thick] (2,2) rectangle (-0.25,0.25);
		\draw [color=black,thick,dashed] (0.5,2) -- (0.5,0.25) (1.25,2) -- (1.25,0.25) (0.5,1.25) -- (-0.25,1.25);
		\node at (0.125,1.625) [font=\tiny]{$-\mathcal{R}^{T}$};
		\node at (1.625,1.125) [font=\tiny]{$-C^{T}_{+}$};
	\end{scope}
	\begin{scope}[xshift=0.0cm,yshift=0.0cm]
		\draw [color=black,thick] (2,-2) rectangle (-0.25,0.25);
		\draw [color=black,thick,dashed] (-0.25,-1.25) -- (2,-1.25) (1.25,0.25) -- (1.25,-2);
		\node at (0.5,-0.5) [font=\Large]{$0$};
		\node at (1.625,-1.625) [font=\Large]{$0$};
		\node at (1.625,-0.5) [font=\Large]{$\mathcal{B}$};
		\node at (0.5,-1.625) [font=\Large]{$-\mathcal{B}^{T}$};
	\end{scope}
	\begin{scope}
		\foreach \x in {2,0.25,-1.25}
			\draw (-2,\x) -- ++ (-0.5,0);
		\draw [stealth-stealth] (-2.25,2) -- (-2.25,0.25) node [pos=0.5,rotate=90,fill=white,inner sep=1.2pt]{$n-2$};
		\draw [stealth-stealth] (-2.25,0.25) -- (-2.25,-1.25) node [pos=0.5,rotate=90,fill=white,inner sep=1.2pt]{$k+2$};
	\end{scope}
	\begin{scope}[xshift=-0.0cm,yshift=0.0cm]
		\draw [color=black,thick,
		] (-2,2) rectangle (-0.25,0.25);
		\node at (-1.125,1.125) [
		inner sep=1pt,font=\Large] {${A}$};
	\end{scope}
	\begin{scope}
		\foreach \x in {-2,-0.25,2}
			\draw (\x,2) -- ++ (0,0.5);
		\draw [stealth-stealth] (-2,2.25) -- (-0.25,2.25) node [pos=0.5,fill=white,inner sep=1.2pt]{$n-2$};
		\draw [stealth-stealth] (-0.25,2.25) -- (2,2.25) node [pos=0.5,fill=white,inner sep=1.2pt]{$n$};
	\end{scope}
\end{tikzpicture}}\quad .
\Label{eq:psishape}
\end{equation}
Next, we carry on to find the elementary transformation that can make the blue shaded region zero in the above equation. According to the definition of rank, the $r$ rows and columns selected by $\mathcal{R}$ correspond to two maximum sets of linear-independent vectors in $C_{-}$. The linear relations with other rows and columns can be solved from the following equations of $x$ and $y$:
{\allowdisplaybreaks
\begin{align}
	&\text{columns:}& &\mathcal{R}\begin{pmatrix}
		x_{i_1,\underline{r+1}} & \cdots & x_{i_1,\underline{n-2}} \\
		\vdots & & \vdots \\
		x_{i_r,\underline{r+1}} & \cdots & x_{i_r,\underline{n-2}} 
	\end{pmatrix}\equiv\mathcal{R}\pmb{x}=\begin{pmatrix}
		(C_{-})_{i_1,\underline{r+1}} & \cdots & (C_{-})_{i_1,\underline{n-2}} \\
		\vdots & & \vdots \\
		(C_{-})_{i_r,\underline{r+1}} & \cdots & (C_{-})_{i_r,\underline{n-2}} \\
	\end{pmatrix}\,,\\
	&\text{rows:}& &\begin{pmatrix}
		y_{\overline{r+1},j_1} & \cdots & y_{\overline{r+1},j_r} \\
		\vdots & & \vdots \\
		y_{\overline{k+2},j_1} & \cdots & y_{\overline{k+2},j_r} \\
	\end{pmatrix}\mathcal{R}\equiv\pmb{y}\mathcal{R}=\begin{pmatrix}
		(C_{-})_{\overline{r+1},j_1} & \cdots & (C_{-})_{\overline{r+1},j_r} \\
		\vdots & & \vdots \\
		(C_{-})_{\overline{k+2},j_1} & \cdots & (C_{-})_{\overline{k+2},j_r} \\
	\end{pmatrix}\,,
\end{align}}
where the underlined and overlined indices are from:
\begin{align*}
	&\{\underline{r+1},\ldots,\underline{n-2}\}=\{1,\ldots,n-2\}\backslash\,\mathsf{j}_{r}\,,& &\{\overline{r+1},\ldots,\overline{k+2}\}=\{1,\ldots,k+2\}\backslash\,\mathsf{i}_{r}\,.
\end{align*}
The matrix $\pmb{x}\equiv(x_{ij})$ is $r\times (n-2-r)$ dimensional and $\pmb{y}\equiv(y_{ij})$ is $(k+2-r)\times r$ dimensional. In particular, the elements of $\pmb{x}$ can be calculated using the Cramer's rule:
\begin{align}
	& x_{i_ka}=\frac{1}{\det({\mathfrak{C}_{-}^{r}})}\frac{[aq]}{[j_kq]}\det\begin{bmatrix}
		j_k \\ \pmb{c}_{a}
	\end{bmatrix}\,, & &i_k\in\mathsf{i}_{r},\quad a\in\{1,2,\ldots,n-2\}\backslash\,\mathsf{j}_{r}\,,
\Label{eq:xia}
\end{align}
where the bracket stands for the matrix obtained from $\mathfrak{C}^{r}_{-}$ by replacing the column $$\pmb{c}_{j_k}\equiv\left(\,(\mathfrak{C}_{-})_{i_1j_k},\ldots,(\mathfrak{C}_{-})_{i_rj_k}\right)^{T}$$ by the column $\pmb{c}_{a}\equiv\left(\,(\mathfrak{C}_{-})_{i_1a},\ldots,(\mathfrak{C}_{-})_{i_ra}\right)^{T}$. More generally, we use
\begin{equation*}
  \begin{bmatrix}
    j_{1} & j_{2} & \cdots \\
    \pmb{c}_{a_1} & \pmb{c}_{a_2} & \cdots \\
  \end{bmatrix}
\end{equation*}
to stand for the matrix obtained from $\mathfrak{C}_{-}^{r}$ by replacing the columns $\{\pmb{c}_{j_1},\pmb{c}_{j_2},\cdots\}$ by $\{\pmb{c}_{a_1},\pmb{c}_{a_2},\cdots\}$. The elements of $\pmb{y}$ have a similar expression, but our calculation does not call for it. From the linear relations, we can write down two elementary transformations:
\begin{align}
	&\pmb{P}_{1}=\begin{pmatrix}
		\pmb{1}_{(n-2)} & & & \\
		 & \pmb{1}_{r} & 0 & & \\
		 & -\pmb{y} & \pmb{1}_{(k+2-r)} & \\
		 & & & \pmb{1}_{(n-k-2)} \\
	\end{pmatrix}\,,& &\pmb{P}_{2}=\begin{pmatrix}
		\pmb{1}_{r} & -\pmb{x} & \\
		0 & \pmb{1}_{n-2-r} & \\
		 & & \pmb{1}_{n} \\
	\end{pmatrix}\,,
\end{align}
where $\pmb{1}_{r}$ is an $r\times r$ identity matrix, for example. They together can make the blue shaded region of \eqref{eq:psishape} zero through the action:
\begin{equation}
\label{eq:psitrans}
	\pmb{\psi}\;\rightarrow\;(\pmb{P}_{2}^{T}\pmb{P}_{1})\pmb{\psi}(\pmb{P}_{1}^{T}\pmb{P}_{2})\,.
\end{equation}
This is what we expect since $\pmb{x}$ and $\pmb{y}$ are deliberately chosen to do the job. However, the surprise comes from the fact that after the action, the lower right $(n-2-r)\times(n-2-r)$ block of ${A}$ also becomes identically zero. In other words, the following equation holds:
\begin{align}
& A_{ab}-\sum_{k=1}^{r}\left(A_{aj_{k}}x_{i_{k}b}+A_{j_{k}b}x_{i_{k}a}\right)+\sum_{k,s=1}^{r}x_{i_{k}a}A_{j_{k}j_{s}}x_{i_{s}b}=0\,,
\Label{eq:Aab}
\end{align}
for all $a,b\in\{1,2,\ldots,n-2\}\backslash\,\mathsf{j}_{r}$.
The proof is given in Appendix~\ref{sec:proof}. Now the matrix $\pmb{\psi}$ becomes:
\begin{equation}
\label{eq:psires1}
\pmb{\psi}=\adjustbox{raise=-2cm}{\begin{tikzpicture}
	\path [pattern=north east lines,pattern color=red!50] (-2,-0.5) -- (-1.25,-0.5) -- (-1.25,1.25) -- (0.5,1.25) -- (0.5,2) -- (-2,2)-- cycle (-2,-2) rectangle (1.25,-1.25) (2,2) rectangle (1.25,-1.25);
	\draw [color=black,thick,dashed] (-1.25,1.25) rectangle (1.25,-1.25) (-1.25,-0.5) -- (-2,-0.5) (-1.25,0.25) -- (-2,0.25) (0.5,1.25) -- (0.5,2) (-0.25,1.25) -- (-0.25,2) (-2,-1.25) -- (-1.25,-1.25) (1.25,2) -- (1.25,1.25) (-0.25,-1.25) -- (-0.25,-2) (1.25,0.25) -- (2,0.25) (1.25,-2) -- (1.25,-1.25) -- (2,-1.25);
	\node at (1.625,-1.625) [font=\Large]{$0$};
	\node at (-1.625,-0.875) [font=\Large]{$0$};
	\node at (0.875,1.625) [font=\Large]{$0$};
	\node at (0,0) [font=\huge]{$0$};
	\draw [color=black,thick] (-2,-2) rectangle (2,2);
	\node at (-1.625,-0.125) [font=\Large,fill=white,inner sep=1pt]{$\mathcal{R}$};
	\node at (0.125,1.625) [font=\tiny,fill=white,inner sep=1pt]{$-\mathcal{R}^{T}$};
	\node at (-1.125,-1.625) [font=\Large,fill=white,inner sep=1pt]{${C}_{+}$};
	\node at (1.625,1.125) [font=\tiny,fill=white,inner sep=1pt]{$-C^{T}_{+}$};
	\node at (1.625,-0.5) [font=\Large,fill=white,inner sep=1pt]{$\mathcal{B}$};
	\node at (0.5,-1.625) [font=\Large,fill=white,inner sep=1pt]{$-\mathcal{B}^{T}$};
	\foreach \x in {2,0.25,-1.25}
	\draw (-2,\x) -- ++ (-0.5,0);
	\draw [stealth-stealth] (-2.25,2) -- (-2.25,0.25) node [pos=0.5,rotate=90,fill=white,inner sep=1.2pt]{$n-2$};
	\draw [stealth-stealth] (-2.25,0.25) -- (-2.25,-1.25) node [pos=0.5,rotate=90,fill=white,inner sep=1.2pt]{$k+2$};
	\begin{scope}
	\foreach \x in {-2,-0.25,2}
	\draw (\x,2) -- ++ (0,0.5);
	\draw [stealth-stealth] (-2,2.25) -- (-0.25,2.25) node [pos=0.5,fill=white,inner sep=1.2pt]{$n-2$};
	\draw [stealth-stealth] (-0.25,2.25) -- (2,2.25) node [pos=0.5,fill=white,inner sep=1.2pt]{$n$};
	\end{scope}
	\end{tikzpicture}}\quad ,
\end{equation}
where only the regions shaded by red are nonzero. 

{
To illustrate these matrix operations, we consider as an example the $6$-point NMHV configuration $(---+++)$ with a solution $\{\sigma\}$ that makes $\rank(C_{-})=2$. According to our convention, $\pmb{\psi}=\Psi_{56}^{56}$. We assume for convenience that the first $2\times 2$ principal minor is nonzero, and choose it as our reference matrix $\mathcal{R}$:
\begin{equation*}
\mathcal{R}=\begin{pmatrix}
(C_-)_{11} & (C_-)_{12} \\
(C_-)_{21} & (C_{-})_{22} \\
\end{pmatrix}\,.
\end{equation*}
Then the transformation matrix $\pmb{P}_{1}$ and $\pmb{P}_{2}$ have the following forms:
\begin{align}
&\pmb{P}_{1}=\left(\begin{array}{ccccc}
\pmb{1}_{4} & & & & \\
& 1 & 0 & 0 & \\
& 0 & 1 & 0 & \\
& -y_{31} & -y_{32} & 1 & \\
& & & & \pmb{1}_{3} \\
\end{array}\right)\,,& &\pmb{P}_{2}=\left(\begin{array}{ccccc}
1 & 0 & -x_{13} & -x_{14} & \\
0 & 1 & -x_{23} & -x_{24} & \\
0 & 0 & 1 & 0 & \\
0 & 0 & 0 & 1 & \\
& & & & \pmb{1}_{6} \\
\end{array}\right)\,.
\end{align}
The matrix elements $y_{31}$ and $y_{32}$ are given by:
\begin{equation}
\renewcommand{\arraystretch}{1.3}
\arraycolsep=0.4em
 y_{31}=\frac{[1q]}{[3q]}\left|\begin{array}{cc}
\frac{\langle 31\rangle}{\sigma_{31}} & \frac{\langle 32\rangle}{\sigma_{32}} \\ 
\frac{\langle 21\rangle}{\sigma_{21}} & -\Sigma_{2}^{-} 
\end{array}\right|\left|\begin{array}{cc}
-\Sigma_{1}^{-} & \frac{\langle 12\rangle}{\sigma_{12}} \\
\frac{\langle 21\rangle}{\sigma_{21}} & -\Sigma_{2}^{-} \\ 
\end{array}\right|^{-1}\,,\qquad y_{32}=\frac{[2q]}{[3q]}\left|\begin{array}{cc}
-\Sigma_{1}^{-} & \frac{\langle 12\rangle}{\sigma_{12}} \\ 
\frac{\langle 31\rangle}{\sigma_{31}} & \frac{\langle 32\rangle}{\sigma_{32}}
\end{array}\right|\left|\begin{array}{cc}
-\Sigma_{1}^{-} & \frac{\langle 12\rangle}{\sigma_{12}} \\
\frac{\langle 21\rangle}{\sigma_{21}} & -\Sigma_{2}^{-} \\ 
\end{array}\right|^{-1}\,.
\end{equation}
The matrix element $x_{1i}$ and $x_{2i}$, with $i=3,4$, are given by:
\begin{equation}
\renewcommand{\arraystretch}{1.3}
\arraycolsep=0.4em
x_{1i}=\frac{[iq]}{[1q]}\left|\begin{array}{cc}
\frac{\langle 1i\rangle}{\sigma_{1i}} & \frac{\langle 12\rangle}{\sigma_{12}} \\
\frac{\langle 2i\rangle}{\sigma_{2i}} & -\Sigma_{2}^{-}
\end{array}\right|\left|\begin{array}{cc}
-\Sigma_{1}^{-} & \frac{\langle 12\rangle}{\sigma_{12}} \\
\frac{\langle 21\rangle}{\sigma_{21}} & -\Sigma_{2}^{-} \\
\end{array}\right|^{-1}\,,\qquad x_{2i}=\frac{[iq]}{[2q]}\left|\begin{array}{cc}
-\Sigma_{1}^{-} & \frac{\langle 1i\rangle}{\sigma_{1i}} \\
\frac{\langle 21\rangle}{\sigma_{21}} & \frac{\langle 2i\rangle}{\sigma_{2i}}
\end{array}\right|\left|\begin{array}{cc}
-\Sigma_{1}^{-} & \frac{\langle 12\rangle}{\sigma_{12}} \\
\frac{\langle 21\rangle}{\sigma_{21}} & -\Sigma_{2}^{-} \\
\end{array}\right|^{-1}\,.
\end{equation}
Then the transformation (\ref{eq:psitrans}) will lead to the form (\ref{eq:psires1}). In particular, we can verify that:
\begin{equation*}
A_{34}-\sum_{j=1}^{2}(A_{3j}x_{j4}+A_{j4}x_{j3})+\sum_{k,s=1}^{2}x_{k3}A_{ks}x_{s4}=0\,,
\end{equation*}
either analytically by the method given in Appendix~\ref{sec:proof}, or just numerically.
}

Finally, using again that $\mathcal{R}$ is full rank, we can make another elementary transformation to make the first $r$ rows and columns of $\pmb{\psi}$ zero except for $\mathcal{R}$ itself, such that: 
\begin{align}\Label{eq:rfresult}
\text{Pf}\,(\pmb{\psi})=\text{Pf}\;\adjustbox{raise=-2cm}{\begin{tikzpicture}
	\path [pattern=north east lines,pattern color=red!50] (-2,-0.5) rectangle (-1.25,0.25) (0.5,2) rectangle (-0.25,1.25) (-1.25,-2) -- (-1.25,-1.25) -- (1.25,-1.25) -- (1.25,1.25) -- (2,1.25) -- (2,-2) -- cycle;
	\draw [color=black,thick,dashed] (-1.25,1.25) rectangle (1.25,-1.25) (-1.25,-0.5) -- (-2,-0.5) (-1.25,0.25) -- (-2,0.25) (0.5,1.25) -- (0.5,2) (-0.25,1.25) -- (-0.25,2) (-1.25,-2) -- (-1.25,-1.25) (2,1.25) -- (1.25,1.25) (-0.25,-1.25) -- (-0.25,-2) (1.25,0.25) -- (2,0.25) (1.25,-2) -- (1.25,-1.25) -- (2,-1.25);
	\draw [color=black,thick] (-2,-2) rectangle (2,2);
	\node at (-1.625,-0.125) [font=\Large,fill=white,inner sep=1pt]{$\mathcal{R}$};
	\node at (0.125,1.625) [font=\tiny,fill=white,inner sep=1pt]{$-\mathcal{R}^{T}$};
	\node at (0,0) [font=\Large]{$0$};
	\end{tikzpicture}}=\det(\mathcal{R})\text{Pf}\;
	\adjustbox{raise=-1.25cm}{\begin{tikzpicture}
	\path [pattern=north east lines,pattern color=red!50] (1.25,1.25) rectangle (0.5,-1.25);
	\path [pattern=north east lines,pattern color=blue!50] (-1.25,-1.25) rectangle (0.5,-0.5);
	\draw [color=black,thick,dashed] (-1.25,-0.5) -- (0.5,-0.5) -- (0.5,1.25) (0.5,-1.25) -- (0.5,-0.5) -- (1.25,-0.5);
	\draw [color=black,thick] (-1.25,-1.25) rectangle (1.25,1.25);
	\draw [color=black,thick,dashed] (-0.25,-1.25) -- (-0.25,-0.5) (1.25,0.25) -- (0.5,0.25);
	\node at (-0.375,0.375) [font=\Large]{$0$};
	\foreach \x in {-1.25,0.5}
		\draw (\x,1.25) -- ++ (0,0.5);
	\foreach \x in {1.25,-1.25}
		\draw (-1.25,\x) -- ++ (-0.5,0);
	\draw [stealth-stealth] (-1.25,1.5) -- (0.5,1.5) node [pos=0.5,fill=white,inner sep=1.2pt,font=\scriptsize]{$n+k-2r$};
	\draw [stealth-stealth] (-1.5,-1.25) -- (-1.5,1.25) node [pos=0.5,fill=white,inner sep=1.2pt,rotate=90,font=\scriptsize]{$2n-2-2r$};
	\end{tikzpicture}}\quad .\\
	(\text{up to a minus sign})\nonumber
\end{align}
Now in the last expression, the lower left block (shaded by blue) has dimension $(n-k-2)\times (n+k-2r)$. For $r\leqslant k$, the columns are more than the rows such that there must exist an elementary transformation that makes one entire column zero, leading to $\text{Pf}\,(\pmb{\psi})=0$. Only at $r=k+1$ (thus $m=k$) do we possibly have a nonzero result:
\begin{equation}
\label{eq:rpfactorize}
	\text{Pf}\,(\pmb{\psi})=\det(\mathcal{R})\det\,\adjustbox{raise=-0.65cm}{\begin{tikzpicture}
		\draw [thick,color=black,pattern=north east lines,pattern color=blue!50] (0,0) rectangle (1.5,1.5);
		\draw (0,1.5) -- ++(0,0.5) (1.5,1.5) -- ++(0,0.5);
		\draw [thick,dashed] (0.8,0) -- (0.8,1.5);
		\draw [stealth-stealth] (0,1.75) -- (1.5,1.75) node [pos=0.5,fill=white,inner sep=1.2pt,font=\tiny]{$n-k-2$};
		\end{tikzpicture}}\qquad(\text{up to a minus sign}) ,
\end{equation} 
which proves part of the Theorem~\ref{thm:1}: \emph{For those solutions with $m\leqslant k$,} $\text{Pf}\,'(\Psi)\neq 0$ \emph{only happens for $m=k$ at} $\text{N}^{k}\text{MHV}$. We note that the blue block in the above two equations consists of both columns from $C_{+}$ and $-\mathcal{B}^{T}$, schematically separated by the dashed line in the middle. Although for $m=k$ the $C_{+}$ part has rank deficiency, the whole matrix is in general still of full rank.

Those solutions with $m>k$ make $C_{-}$ be of full rank and $\rank(C_{+})=r'\leqslant n-k-4$, according to \eqref{eq:rankCpm}. We can move $C_{+}$ above $C_{-}$ and then perform the same calculation to the $C_{+}$ part, exchanging all angular and square brackets involved in the above calculation. At the end, we can still reach \eqref{eq:rfresult}, while the dimension of the upper left zero block is now $$(2n-k-2r'-4)\times(2n-k-2r'-4)\,.$$ 
Then blue block in \eqref{eq:rfresult} has dimension:
\begin{equation*}
(k+2)\times(2n-k-2r'-4)\,.
\end{equation*}
It always has more columns than rows when $r'\leqslant n-k-4$, which leads to $\text{Pf}\,(\pmb{\psi})=0$. Therefore, we have proved that \emph{at} $\text{N}^{k}\text{MHV}$, \emph{all those solutions with $m>k$ lead to} $\text{Pf}\,'(\Psi)=0$. Now we have completed the proof that under the partition $\pmb{P}_{-}$, $\text{Pf}\,'(\Psi)=0$ for all $m$ except for $m=k$. Finally, we note that the $\pmb{P}_{+}$ and $\text{N}^{n-k-4}\text{MHV}$ part can be proved identically, which completes our proof of Theorem~\ref{thm:1}.

\section{Support of general EYM amplitudes on the solutions}
\label{sec:generalamp}
As already shown in \eqref{eq:theories}, the EYM amplitudes in the CHY formalism also contains $\text{Pf}\,'(\Psi)$ so that our Theorem~\ref{thm:1} can apply to EYM. On the other hand, the interplay between the factor $\text{Pf}\,(\Psi_{\mathsf{h}})$ and the solutions enables us to prove a previously raised conjecture~\cite{Bern:1999bx}: \emph{the EYM amplitudes with gluons having the same helicities must vanish.}

Because of the similarity between the matrix $\Psi$ and $\Psi_{\mathsf{h}}$, the method developed in the previous section can also apply here.
Moreover, since Theorem~\ref{thm:1} has now been proved, we can use the corollary~(\ref{eq:rankCpm2}) here. 

Suppose we are now looking at a solution $\{\sigma\}\in\pmb{P}_{-}(n-3,m)$, such that:
\begin{align}
&\rank[(C_{\mathsf{h}})_{-}]=\min\{m+1,s^{-}\}\equiv r\,,& &\rank[(C_{\mathsf{h}})_{+}]=\min\{n-m-3,s^{+}\}\equiv r'\,.
\end{align}
First, in $(C_{\mathsf{h}})_{-}$ we can choose a $r\times r$ submatrix $\mathcal{R}_{\mathsf{h}}$ as reference. Following the same procedure as the previous section, we can reach at:
\begin{align}
\text{Pf}\,(\Psi_{\mathsf{h}})=\det(\mathcal{R}_{\mathsf{h}})\text{Pf}\;
\adjustbox{raise=-1.25cm}{\begin{tikzpicture}[scale=1.1]
	\path [pattern=north east lines,pattern color=red!50] (1.25,1.25) rectangle (0.5,-1.25);
	\path [pattern=north east lines,pattern color=blue!50] (-1.25,-1.25) rectangle (0.5,-0.5);
	\draw [color=black,thick,dashed] (-1.25,-0.5) -- (0.5,-0.5) -- (0.5,1.25) (0.5,-1.25) -- (0.5,-0.5) -- (1.25,-0.5);
	\draw [color=black,thick] (-1.25,-1.25) rectangle (1.25,1.25);
	\draw [color=black,thick,dashed] (-0.25,-1.25) -- (-0.25,-0.5) (1.25,0.25) -- (0.5,0.25);
	\node at (-0.375,0.375) [font=\Large]{$0$};
	\foreach \x in {-1.25,0.5}
	\draw (\x,1.25) -- ++ (0,0.5);
	\foreach \x in {1.25,-1.25}
	\draw (-1.25,\x) -- ++ (-0.5,0);
	\draw [stealth-stealth] (-1.25,1.5) -- (0.5,1.5) node [pos=0.5,fill=white,inner sep=1.2pt,font=\scriptsize]{$s+s^{-}-2r$};
	\draw [stealth-stealth] (-1.5,-1.25) -- (-1.5,1.25) node [pos=0.5,fill=white,inner sep=1.2pt,rotate=90,font=\scriptsize]{$2s-2r$};
	\end{tikzpicture}}\quad(\text{up to a minus sign})\,.
\end{align}
This Pfaffian vanishes if the lower left block has more column than rows, which happens if $r\leqslant s^{-}-1$. It means that if $(C_{\mathsf{h}})_{-}$ has a rank deficiency, $\text{Pf}\,(\Psi_{\mathsf{h}})=0$. If we perform the same transformation on the $(C_{\mathsf{h}})_{+}$ part, we may find that $\text{Pf}(\Psi_{\mathsf{h}})=0$ also when $r'\leqslant s^{+}-1$. Therefore, our conclusion is thus:
\begin{equation}
\text{Pf}\,(\Psi_{\mathsf{h}})=0\quad\text{if}\quad m\leqslant s^{-}-2\quad\text{or}\quad m\geqslant n-s^{+}-2\,.
\end{equation}
Depending on the numbers of positive and negative helicity gravitons, the support of $\text{Pf}(\Psi_{\mathsf{h}})$ is thus:
\begin{equation}
\bigcup_{m=s^{-}-1}^{n-3-s^{+}}\pmb{P}_{-}(n-3,m)\,.
\Label{eq:generalpattern}
\end{equation}
In this expression, we have implicitly defined that $\pmb{P}_{-}(n-3,-1)=\pmb{P}_{-}(n-3,n-3)=\varnothing$, which corresponds to $s^{\pm}=0$ respectively. {The physical meaning of \eqref{eq:generalpattern} is that: with the graviton helicity configuration $(s^{-},s^{+})$ fixed, the only possible nonzero $\text{N}^{k}\text{MHV}$ amplitudes are restricted to:
\begin{equation}
	\max\{\,0,\,s^{-}-1\,\}\leqslant k\leqslant\min\{\,n-4,\,n-3-s^{+}\,\}\,.
\Label{eq:krange}
\end{equation}}

Knowing the support of $\text{Pf}(\Psi_{\mathsf{h}})$, we can now prove that the same-gluon-helicity amplitudes of EYM mush vanish. At $\text{N}^{k}\text{MHV}$, if all the gluons have positive helicity, then the number of negative gravitons must be:
\begin{equation*}
s^{-}=k+2\,,
\end{equation*}
such that 
\begin{equation*}
\text{Pf}\,(\Psi_{\mathsf{h}})\neq 0\quad\text{only on the solution set}\quad\bigcup_{m=k+1}^{n-3-s^{+}}\pmb{P}_{-}(n-3,m)\quad\text{if gluons are all-plus.}
\end{equation*}
Since the support of $\text{Pf}\,'(\Psi)$, $\pmb{P}_{-}(n-3,k)$, is not contained in the support of $\text{Pf}\,(\Psi_{\mathsf{h}})$, this amplitude must vanish. Similarly, if all the gluons have negative helicity, then the number of positive gravitons must be:
\begin{equation*}
s^{+}=n-k-2\,,
\end{equation*}
such that
\begin{equation*}
\text{Pf}\,(\Psi_{\mathsf{h}})\neq 0\quad\text{only on the solution set}\quad\bigcup_{m=s^{-}-1}^{k-1}\pmb{P}_{-}(n-3,m)\quad\text{if gluons are all-minus.}
\end{equation*}
The set $\pmb{P}_{-}(n-3,k)$ is still not included. As a result the amplitude still vanishes. We have thus completed our proof that \emph{the tree-level single-trace EYM amplitudes with gluons having the same helicities must vanish.} Together with \eqref{eq:krange}, we know that nonzero amplitudes contain at least one gluon with a different helicity from the others.

\section{Numerical study}\label{sec:numerical}
In this section, we provide a numerical study verifying our analytical results. More importantly, it reveals that the rank characterization established in Sec.~\ref{sec:chara} leads to an Eulerian number pattern for $\text{N}^{k}\text{MHV}$ solutions.
\subsection{General pattern}
We have conducted a numerical study up to $n=8$ with all helicity configurations, and checked the value of $\text{rank}(\mathfrak{C}_{\pm})$ for each solution. The result agrees with \eqref{eq:rankCpm2}. One thing that remains to be investigated is to count how many solutions fall into each subset of our partition $\pmb{P}_{\pm}$. The numerical study results in the following pattern:
\paragraph{Numerical Fact:}	Both the partition $\pmb{P}_{-}$ and $\pmb{P}_{+}$ have an Eulerian number pattern:
	\begin{equation}
	|\pmb{P}_{\pm}(n-3,m)|=A(n-3,m)\,.
	\end{equation}
This conjecture, together with Theorem~\ref{thm:1}, exactly reproduce the statement that only $A(n-3,k)$ solutions support $\text{N}^{k}\text{MHV}$ amplitudes~\cite{Cachazo:2013iaa,Cachazo:2016sdc}. The rank condition proposed above gives a set of extra equations satisfied only by those solutions in a certain subset. For example, it will be interesting to study why by associating the following equations: 
\begin{align}
&\left|\begin{array}{cccc}
(\mathfrak{C}_{-})_{i_{1}j_{1}} & (\mathfrak{C}_{-})_{i_{1}j_{2}} & \cdots & (\mathfrak{C}_{-})_{i_{1}j_{r}} \\
\vdots & \vdots & & \vdots \\
(\mathfrak{C}_{-})_{i_{r}j_{1}} & (\mathfrak{C}_{-})_{i_{r}j_{2}} & \cdots & (\mathfrak{C}_{-})_{i_{r}j_{r}} \\
\end{array}\right|=0\,, & &\forall\{i_{1},\ldots,i_{r}\},\{j_{1},\ldots,j_{r}\}\subset\mathsf{p}
\Label{eq:rank}
\end{align}
with the scattering equations (\ref{eq:SE}), we can produce the solution set:
\begin{equation*}
\bigcup_{m=0}^{r-1}\pmb{P}_{-}(n-3,m)\,,
\end{equation*}
whose order is supposed to be $\sum_{m=0}^{r-1}A(n-3,m)$. It will be our future work to study the consistency condition between \eqref{eq:rank} and (\ref{eq:SE}) to reveal why this seemingly over-determined set of equations can still produce the proclaimed number of solutions. 

\subsection{\texorpdfstring{$n=7$}{n=7} and \texorpdfstring{$n=8$}{n=8} examples}
\begin{table}
	\renewcommand{\arraystretch}{1.5}
	\centering
	\begin{tabularx}{\textwidth}{|c|*{9}{C|}}
		\hline
		& \multicolumn{4}{c|}{$n=7$} & \multicolumn{5}{c|}{$n=8$} \\ \hline
		$\text{N}^{k}\text{MHV}$                          & $0$ & $1$ & $2$ & $3$ & $0$ & $1$ & $2$ & $3$ & $4$ \\ \hline
		$\text{Pf}\,'(\Psi)\neq 0$ & $\pmb{1}$   & $\pmb{11}$   & $\overline{\pmb{11}}$ & $\overline{\pmb{1}}$ & $\pmb{1}$ & $\pmb{26}$ & $\pmb{66}$ & $\overline{\pmb{26}}$ & $\overline{\pmb{1}}$ \\ \hline
		$\rank(\mathfrak{C}_{-})$ & $1$ & $2$ & $3$ & $4$ & $1$ & $2$ & $3$ & $4$ & $5$ \\ \hline
                $\rank(\mathfrak{C}_{+})$ & $4$ & $3$ & $2$ & $1$ & $5$ & $4$ & $3$ & $2$ & $1$ \\ \hline
	\end{tabularx}
	\caption{\label{tab:PfPsi} The solution sets that support $\text{Pf}\,'(\Psi)$ at different helicity configurations at $n=7$ and $n=8$. For example, the solutions in $\overline{\pmb{11}}$ are complex conjugate to those in $\pmb{11}$, and support $\overline{\text{NMHV}}$ instead of NMHV. }
\end{table}
At $n=7$ and $n=8$, the Eulerian numbers $A(n-3,k)$ are given by:
\begin{equation}
	\setlength{\columnwidth}{10pt}
	\begin{array}{*{7}{>{\centering $} m{1.5cm} <{$}}}
			& (n-3)! & k=0 & k=1 & k=2 & k=3 & k=4 \tabularnewline \hline\hline
		n=7 & 24 & 1  & 11  & 11  & 1 & - \tabularnewline
		n=8 & 120 & 1 & 26 & 66 & 26 & 1 \tabularnewline \hline
	\end{array}
\end{equation}
while the pattern in the solutions that support $\text{Pf}\,'(\Psi)$ is shown in Tab.~\ref{tab:PfPsi}. Here we simplify our notation by denoting the solution subset that supports $n$-point $\text{N}^{k}\text{MHV}$ amplitudes by its order in bold face. The solution sets with a bar are complex conjugate to the unbarred ones, which support the amplitudes with the helicities flipped. Tab.~\ref{tab:PfPsi} exactly exhibits the pattern stated in the previous subsection: \emph{at} $\text{N}^{k}\text{MHV}$, $\text{Pf}\,'(\Psi)$ \emph{is supported only by the subset $\pmb{P}_{-}(n-3,k)$, whose order is the Eulerian number $A(n-3,k)$.}

In contrast, the solution subset that supports $\text{Pf}\,(\Psi_{\mathsf{h}})$ in EYM exhibit a summation pattern, as shown in Tab.~\ref{tab:PfPsih}, which also agrees exactly with \eqref{eq:generalpattern}. {For example, at $n=8$ and $(s^{-},s^{+})=(2,2)$, the support of $\text{Pf}\,(\Psi_{\mathsf{h}})$ is $\pmb{26}+\pmb{66}+\overline{\pmb{26}}$, which indicates that the nonzero $\text{N}^{k}\text{MHV}$ amplitudes are NMHV, $\text{N}^{2}\text{MHV}$ and $\overline{\text{NMHV}}$. The same-gluon-helicity amplitudes, MHV and $\overline{\text{MHV}}$, vanish since their supports, $\pmb{1}$ and $\overline{\pmb{1}}$, are not included in the support of $\text{Pf}\,(\Psi_{\mathsf{h}})$.} At $s=0$, by definition $\text{Pf}\,(\Psi_{\mathsf{h}})=1$ and we return to tree-level Yang-Mills amplitudes. 
\begin{table}
	\renewcommand{\arraystretch}{1.3}
	\centering
	\begin{tabular}{|c|*{7}{c|}}
		\cline{1-7}
		& \multicolumn{6}{c|}{$n=7$, $\text{Pf}\,(\Psi_{\mathsf{h}})\neq 0$} & \multicolumn{1}{c}{}\\ \cline{2-7}
		& $s^{-}=0$  & $s^{-}=1$  & $s^{-}=2$ & $s^{-}=3$          & $s^{-}=4$ & $s^{-}=5$ & \multicolumn{1}{c}{}\\ \cline{1-7}
		$s^{+}=0$ & $\pmb{24}$ & $\pmb{24}$ & $\pmb{11}+\overline{\pmb{11}}+\overline{\pmb{1}}$ & $\overline{\pmb{11}}+\overline{\pmb{1}}$ & $\overline{\pmb{1}}$ & $\varnothing$ & \multicolumn{1}{c}{}\\ \cline{1-7}
		$s^{+}=1$ & $\pmb{24}$ & $\pmb{24}$ & $\pmb{11}+\overline{\pmb{11}}+\overline{\pmb{1}}$ & $\overline{\pmb{11}}+\overline{\pmb{1}}$ & $\overline{\pmb{1}}$ & \multicolumn{2}{c}{} \\ \cline{1-6}
		$s^{+}=2$ & $\pmb{1}+\pmb{11}+\overline{\pmb{11}}$ & $\pmb{1}+\pmb{11}+\overline{\pmb{11}}$ & $\pmb{11}+\overline{\pmb{11}}$ & $\overline{\pmb{11}}$ & \multicolumn{3}{c}{} \\ \cline{1-5}
		$s^{+}=3$ & $\pmb{1}+\pmb{11}$ & $\pmb{1}+\pmb{11}$ & $\pmb{11}$ & \multicolumn{4}{c}{} \\ \cline{1-4}
		$s^{+}=4$ & $\pmb{1}$ & $\pmb{1}$ & \multicolumn{5}{c}{} \\ \cline{1-3}
		$s^{+}=5$ & $\varnothing$ & \multicolumn{6}{c}{} \\ \hline
		& \multicolumn{7}{c|}{$n=8$, $\text{Pf}\,(\Psi_{\mathsf{h}})\neq 0$} \\ \cline{2-8}
		& $s^{-}=0$  & $s^{-}=1$  & $s^{-}=2$ & $s^{-}=3$          & $s^{-}=4$ & $s^{-}=5$ & $s^{-}=6$ \\ \hline
		\multirow{2}{*}{$s^{+}=0$} & \multirow{2}{*}{$\pmb{120}$} & \multirow{2}{*}{$\pmb{120}$} & $\pmb{26}+\pmb{66}$ & $\pmb{66}$ & \multirow{2}{*}{$\overline{\pmb{26}}+\overline{\pmb{1}}$} & \multirow{2}{*}{$\overline{\pmb{1}}$} & \multirow{2}{*}{$\varnothing$} \\
		&			   &			&	${}+\overline{\pmb{26}}+\overline{\pmb{1}}$ & ${}+\overline{\pmb{26}}+\overline{\pmb{1}}$  & & & \\ \hline
		\multirow{2}{*}{$s^{+}=1$} & \multirow{2}{*}{$\pmb{120}$} & \multirow{2}{*}{$\pmb{120}$} & $\pmb{26}+\pmb{66}$ & $\pmb{66}$ & \multirow{2}{*}{$\overline{\pmb{26}}+\overline{\pmb{1}}$} & \multirow{2}{*}{$\overline{\pmb{1}}$} & \multicolumn{1}{c}{} \\
		&			   &	    	& ${}+\overline{\pmb{26}}+\overline{\pmb{1}}$ & ${}+\overline{\pmb{26}}+\overline{\pmb{1}}$ & & & \multicolumn{1}{c}{}\\ \cline{1-7}
		\multirow{2}{*}{$s^{+}=2$} & $\pmb{1}+\pmb{26}$ & $\pmb{1}+\pmb{26}$ & $\pmb{26}$ & \multirow{2}{*}{$\pmb{66}+\overline{\pmb{26}}$} & \multirow{2}{*}{$\overline{\pmb{26}}$} & \multicolumn{2}{c}{} \\
		& ${}+\pmb{66}+\overline{\pmb{26}}$ & ${}+\pmb{66}+\overline{\pmb{26}}$ & ${}+\pmb{66}+\overline{\pmb{26}}$ & & & \multicolumn{2}{c}{} \\ \cline{1-6}
		$s^{+}=3$ & $\pmb{1}+\pmb{26}+\pmb{66}$ & $\pmb{1}+\pmb{26}+\pmb{66}$ & $\pmb{26}+\pmb{66}$ & $\pmb{66}$ & \multicolumn{3}{c}{} \\ \cline{1-5}
		$s^{+}=4$ & $\pmb{1}+\pmb{26}$ & $\pmb{1}+\pmb{26}$ & $\pmb{26}$ & \multicolumn{4}{c}{} \\ \cline{1-4}
		$s^{+}=5$ & $\pmb{1}$ & $\pmb{1}$ & \multicolumn{5}{c}{} \\ \cline{1-3}
		$s^{+}=6$ & $\varnothing$ & \multicolumn{6}{c}{} \\ \cline{1-2}
	\end{tabular}
	\caption{\label{tab:PfPsih} The solution sets that support $\text{Pf}\,(\Psi_{\mathsf{h}})$ at different graviton helicity configurations at $n=7$ and $n=8$. The $s^{+}$ and $s^{-}$ are the numbers of positive and negative helicity gravitons. The amplitude vanishes if the graviton number is more than $n-2$.}
\end{table}

\section{Discussion}\label{sec:dis}
As can be indicated in \eqref{eq:rpfactorize}, the reduced Pfaffian in the \emph{Yang-Mills} integrand factorizes into the product of two determinants after our transformation, one of which is just our discriminant matrix $\mathfrak{C}_{-}$. This factorized integrand has appeared in the twistor and ambitwistor formalism of \emph{gravity amplitudes}~\cite{Cachazo:2012pz,Cachazo:2012kg,Geyer:2014fka}, while the matrices involved there are different but can be related to ours~\cite{Song}. In this section, we will show briefly that the $\Phi$ and $\W{\Phi}$ matrix defined in the gravity integrand of~\cite{Cachazo:2012pz,Cachazo:2012kg} are equivalent to ours after a wise gauge choice.
        
Our discriminant matrix $\mathfrak{C}_{\pm}$ indeed has a close relation to the matrix $\Phi$ and $\W{\Phi}$ in the integrand for $\mathcal{N}=8$ supergravity amplitudes in a twistor-string approach~\cite{Cachazo:2012kg,Cachazo:2012pz}\footnote{We are indebted to Freddy Cachazo for pointing out to us this connection.}. With our notations, the matrix $\W{\Phi}$ is defined as:\footnote{Our use of $[...]$ and $\langle ...\rangle$ is opposite to that in~\cite{Cachazo:2012pz}. In addition, we have used inhomogeneous world sheet coordinates in the expression, and factored out $t_{i}^{2}$ in each row.}
\begin{align}
  & \W{\Phi}_{ij}=\frac{\langle ij\rangle}{\sigma_{ij}}\frac{t_{j}}{t_{i}}\quad (i\neq j)\,,& & \W{\Phi}_{ii}=\sum_{j\neq i}^{n}\frac{\langle ij\rangle}{\sigma_{ij}}\frac{t_{j}\prod_{a=0}^{k+1}\sigma_{jp_a}}{t_{i}\prod_{a=0}^{k+1}\sigma_{ip_a}}\,,
\end{align}
where $p_a$ are arbitrary reference points on the world sheet, and $t_{i}$ is the scaling factor of the world sheet homogeneous coordinates. Ref.~\cite{Cachazo:2012kg,Cachazo:2012pz} proposed that the supergravity amplitude can be obtained by integrating over the supertwistor $Z_{a}^{I}=(\lambda_{a}^{\alpha},\mu_{a}^{\dot{\alpha}},\eta_{a}^{I})$ for $a$ ranging from $0$ to $k+1$ and $I$ from $0$ to $8$. For each external particle, there is a curve in the supertwistor space, described by the degree $k+1$ polynomial:
\begin{equation*}
	Z_{j}^{I}(\sigma_{j})=t_{j}\sum_{a=0}^{k+1}Z_{a}^{I}\sigma_{j}^{a}\,.
\end{equation*}
In particular, for an arbitrary spinor $|q]$, the inner produc{\blue t} $[\lambda_{j}(\sigma_{j})q]$ is a degree $k+1$ polynomial. Then by an appropriate choice of $p_{a}$, we can always make 
\begin{equation*}
[\lambda_{j}(\sigma_{j})q]=t_{j}\prod_{a=0}^{k+1}\sigma_{jp_{a}}\,.
\end{equation*}
Finally, by constraining $\lambda_{j}(\sigma_{j})$ onto the external momentum data by a group of delta functions (which is now known to be equivalent to the scattering equations),  one may find that the diagonal elements $\W{\Phi}_{ii}$  become those of our matrix $\mathfrak{C}_{-}$, which is of a very neat form (\ref{eq:Cpm}).

\section{Conclusion}\label{sec:con}
In this paper, we have studied the relationship between the solutions of scattering equations and the $\text{N}^{k}\text{MHV}$ amplitudes in quantum gauge/gravity theories, in particular in four dmensions. This is a crucial for understanding how the CHY integrand for amplitudes encodes dynamic information of quantum gauge/gravity theories. 
	
In more details, we have defined two discriminant matrices $\mathfrak{C}_{\pm}$, given in \eqref{eq:Cpm}, for a solution to scattering equations. Their ranks satisfy a constraint: the sum of their ranks is equal to $n-2$ for $n$ scattering particles. Either rank gives rise to a partition of the solution set of scattering equations into a disjoint union of subsets. Moreover, those solutions in the subset labeled by $$\rank(\mathfrak{C}_{-})=k+1$$ can only support the $\text{N}^{k}\text{MHV}$ amplitudes. In this way, we have refined our understanding of the solutions by a characterization in terms of their discriminant matrices, and that of the interplay between the solutions and helicity configurations of amplitudes, manifested by the correspondence between the rank classification of solutions and the $\text{N}^{k}\text{MHV}$ (or helicity configuration) classification of amplitudes. 

This rank characterization classifies solutions by a set of algebraic equations consistent with the scattering equations. It will be interesting to further study why the number of solutions that satisfy both our new sets of algebraic equations and the scattering equations equals to an Eulerian number.
	
We note that since the discriminant matrices are related, by \eqref{eq:C-relation}, to the $C$ matrix in the CHY integrand for the $\Psi$ matrix in \eqref{eq:Psi} and (\ref{eq:C}), the above statements in terms of the rank of the $\mathfrak{C}_{\pm}$ matrices can be reformulated in terms of $\rank(C_{\pm})$. The conceptual advantage of using $\mathfrak{C}_{\pm}$ instead of $C_{\pm}$ is that the former depends only on particle kinematics, and is (gauge) independent of the choice of the reference spinors in polarization vectors.

Moreover, using the above technique, we have been able to prove 
analytically that if the gluons in the tree-level single-trace EYM amplitudes have the same helicity, the amplitudes must vanish identically. We have also done numerical verification for $n=7$ and $8$. In addition to the above correspondence, our numerics revealed an Eulerian number pattern, $A(n-3,k)$, for the number of solutions in the subset that supports (or contributes to) the $\text{N}^{k}\text{MHV}$ amplitudes, which agrees with previous claims~\cite{Cachazo:2013iaa,Cachazo:2016sdc}. 
	
In summary, we have shown that in gauge or gravity theory in four dimensions, the solution set of scattering equations is decomposed into disjoint union of subsets, labeled by the rank of their discriminant matrices, each of which supports the $\text{N}^{k}\text{MHV}$ amplitudes with a specific value of $k$. {It will be interesting to study how this pattern generalizes in higher dimensions. We hope that our rank characterization will be useful along this direction.} Also it is believed that this rank characterization may help further understand the $\text{N}^{k}\text{MHV}$ solutions to scattering equations.

\acknowledgments
The authors thank Freddy Cachazo and Song He for very helpful comments. YD would like to acknowledge National Natural Science Foundation of 
China under Grant Nos. 11105118, 111547310, as well as the 351 program of Wuhan University. 

\appendix
\section{Proof of \eqref{eq:Aab}}\label{sec:proof}
The strategy of our proof is to group all the matrix elements involved into $(r+1)\times(r+1)$ minors of $\mathfrak{C}_{-}$, which should vanish identically since we have assumed that the solution under consideration makes $\mathfrak{C}_{-}$ be of rank $r$. 

The last term of \eqref{eq:Aab} can be written as:
\begin{equation*}
\sum_{k,s=1}^{r}x_{i_ka}A_{j_kj_s}x_{i_sb}=\frac{1}{2}\sum_{k,s=1}^{r}A_{j_kj_s}(x_{i_ka}x_{i_sb}-x_{i_sa}x_{i_kb})\,,
\end{equation*}
since $A_{j_kj_s}$ is anti-symmetric. Then we plug in \eqref{eq:xia} for $x$, such that:
\begin{align}
&\quad\frac{[j_kq][j_sq]}{[aq][bq]}[\det(\mathfrak{C}_{-}^{r})]^{2}(x_{i_ka}x_{i_sb}-x_{i_sa}x_{i_kb})\nonumber\\
&=\det\begin{bmatrix}
j_k \\ \pmb{c}_{a}
\end{bmatrix}\det\begin{bmatrix}
j_s \\ \pmb{c}_{b}
\end{bmatrix}-\det\begin{bmatrix}
j_s \\ \pmb{c}_{a}
\end{bmatrix}\det\begin{bmatrix}
j_k \\ \pmb{c}_{b}
\end{bmatrix}\nonumber\\
&=\det\begin{bmatrix}
j_k & j_s \\
\pmb{c}_{a} & \pmb{c}_{j_s} \\
\end{bmatrix}\det\begin{bmatrix}
j_k & j_s \\
\pmb{c}_{j_k} & \pmb{c}_{b} \\
\end{bmatrix}+\det\begin{bmatrix}
j_k & j_s \\
\pmb{c}_{a} & \pmb{c}_{j_k} \\
\end{bmatrix}\det\begin{bmatrix}
j_k & j_s \\
\pmb{c}_{b} & \pmb{c}_{j_s} \\
\end{bmatrix}\nonumber\\
&=\det(\mathfrak{C}_{-}^{r})\det\begin{bmatrix}
j_k & j_s \\
\pmb{c}_{a} & \pmb{c}_{b} \\
\end{bmatrix}\,,
\end{align}
where the sign change in the second equation is due to the exchange of column $\pmb{c}_a$ and $\pmb{c}_{j_k}$: 
\begin{align}
\begin{bmatrix}
j_s \\ \pmb{c}_a
\end{bmatrix}=\begin{bmatrix}
j_s & j_k \\
\pmb{c}_{a} & \pmb{c}_{j_k} \\
\end{bmatrix}=-\begin{bmatrix}
j_k & j_s \\
\pmb{c}_{a} & \pmb{c}_{j_k} \\
\end{bmatrix}\,.
\end{align}
The last bracket means that in the matrix $\mathfrak{C}_{-}^{r}$, the column
\begin{equation*}
\pmb{c}_{j_k}=\left(\,(\mathfrak{C}_{-})_{i_1j_k},\ldots,(\mathfrak{C}_{-})_{i_rj_k}\right)^{T}\quad\text{and}\quad\pmb{c}_{j_s}=\left(\,(\mathfrak{C}_{-})_{i_1j_s},\ldots,(\mathfrak{C}_{-})_{i_rj_s}\right)^{T}
\end{equation*}
are replaced by $\pmb{c}_{a}$ and $\pmb{c}_{j_k}$ respectively. To further simplify the notation, we will use the symbol $\mathfrak{C}$ in favor of $\mathfrak{C}_{-}$.\footnote{For $m>k$ (see the last paragraph of Sec.~\ref{sec:pfpsi}), $\mathfrak{C}$ can also stand for $\mathfrak{C}_{+}$. The derivation is still valid once we replace all square brackets to angular ones, and vice versa.} If we also use the fact that:
\begin{equation}
	A_{ab}=-[ab]\mathfrak{C}_{ab}\,,
\end{equation}
we can transform \eqref{eq:Aab} into:
\begin{align}
	A_{ab}-\sum_{k=1}^{r}\left(A_{aj_{k}}x_{i_{k}b}+A_{j_{k}b}x_{i_{k}a}\right)+\sum_{k,s=1}^{r}x_{i_{k}a}A_{j_{k}j_{s}}x_{i_{s}b}
	=-\frac{1}{\det(\mathfrak{C}^{r})}\mathcal{A}_{ab}\,,
\end{align}
namely, we extract all common factors such that we only need to study:
\begin{align}
\mathcal{A}_{ab}&=[ab]\mathfrak{C}_{ab}\det(\mathfrak{C}^{r})-\sum_{k=1}^{r}\frac{[aj_k][bq]}{[j_kq]}\mathfrak{C}_{aj_k}\det\begin{bmatrix}
j_k \\ \pmb{c}_{b} \\
\end{bmatrix}-\sum_{k=1}^{r}\frac{[j_kb][aq]}{[j_kq]}\mathfrak{C}_{j_kb}\det\begin{bmatrix}
j_k \\ \pmb{c}_{a} \\
\end{bmatrix}\nonumber\\
&\quad +\frac{1}{2}\sum_{k,s=1}^{r}\frac{[j_kj_s][aq][bq]}{[j_kq][j_sq]}\mathfrak{C}_{j_kj_s}\det\begin{bmatrix}
j_k & j_s \\
\pmb{c}_{a} & \pmb{c}_{b} \\
\end{bmatrix}\nonumber\\
&=[ab]\left(\mathfrak{C}_{ab}\det(\mathfrak{C}^{r})-\sum_{k=1}^{r}\mathfrak{C}_{aj_{k}}\det\begin{bmatrix}
	j_{k} \\ \pmb{c}_{b} \\
\end{bmatrix}\right)\nonumber\\
&\quad+\sum_{k=1}^{r}\frac{[aq][j_{k}b]}{[j_{k}q]}\left(\mathfrak{C}_{aj_{k}}\det\begin{bmatrix}
	j_{k} \\ \pmb{c}_{b} \\
\end{bmatrix}-\mathfrak{C}_{j_{k}b}\begin{bmatrix}
	j_{k} \\ \pmb{c}_{a} \\
\end{bmatrix}\right)+\frac{1}{2}\sum_{k,s=1}^{r}\frac{[j_{k}j_{s}][aq][bq]}{[j_{k}q][j_{s}q]}\mathfrak{C}_{j_{k}j_{s}}\det\begin{bmatrix}
	j_{k} & j_{s} \\
	\pmb{c}_{a} & \pmb{c}_{b} \\
\end{bmatrix}\,.
\end{align}
The first line forms an $(r+1)\times(r+1)$ minor:
\begin{equation}
	\left(\mathfrak{C}_{ab}\det(\mathfrak{C}^{r})-\sum_{k=1}^{r}\mathfrak{C}_{aj_{k}}\det\begin{bmatrix}
	j_{k} \\ \pmb{c}_{b} \\
	\end{bmatrix}\right)=\det\left(\begin{array}{c|ccc}
		\mathfrak{C}_{ab} & \mathfrak{C}_{aj_{1}} & \cdots & \mathfrak{C}_{aj_{r}} \\ \hline
		\mathfrak{C}_{i_{1}b} & & & \\
		\vdots & & \mathfrak{C}^{r} & \\
		\mathfrak{C}_{i_{r}b} & & & \\
	\end{array}\right)=0\,,
\end{equation}
which vanishes due to the rank $r$ condition. 
Finally, we need to prove the vanishing of the second line:
\begin{equation}
\sum_{k=1}^{r}\frac{[aq][j_{k}b]}{[j_{k}q]}\left(\mathfrak{C}_{aj_{k}}\det\begin{bmatrix}
j_{k} \\ \pmb{c}_{b} \\
\end{bmatrix}-\mathfrak{C}_{j_{k}b}\begin{bmatrix}
j_{k} \\ \pmb{c}_{a} \\
\end{bmatrix}\right)+\frac{1}{2}\sum_{k,s=1}^{r}\frac{[j_{k}j_{s}][aq][bq]}{[j_{k}q][j_{s}q]}\mathfrak{C}_{j_{k}j_{s}}\det\begin{bmatrix}
j_{k} & j_{s} \\
\pmb{c}_{a} & \pmb{c}_{b} \\
\end{bmatrix}=0\,.
\Label{eq:id2}
\end{equation}
All the matrix elements involved can be packed into the following $(r+1)\times(r+1)$ minor:
\begin{equation}
\det(\mathscr{C}_{k})=\det\begin{pmatrix}
\mathfrak{C}_{i_1j_1} & \cdots &\widehat{\mathfrak{C}}_{i_1j_k} & \cdots & \mathfrak{C}_{i_1j_r} & \mathfrak{C}_{i_1a} & \mathfrak{C}_{i_1b} \\
\vdots & & \vdots & & \vdots & \vdots & \vdots \\
\mathfrak{C}_{i_rj_1} & \cdots &\widehat{\mathfrak{C}}_{i_rj_k} & \cdots & \mathfrak{C}_{i_rj_r} & \mathfrak{C}_{i_ra} & \mathfrak{C}_{i_rb} \\
\mathfrak{C}_{j_{k}j_{1}} & \cdots &\widehat{\mathfrak{C}}_{j_kj_k} & \cdots & \mathfrak{C}_{j_{k}j_{r}} & \mathfrak{C}_{j_ka} & \mathfrak{C}_{j_kb} \\
\end{pmatrix}=0\,,
\end{equation}
where a hat means that this element is deleted. After a proper arrangement, the expansion along the last row of $\mathscr{C}_{k}$ gives:
\begin{align}
\left(\mathfrak{C}_{aj_{k}}\det\begin{bmatrix}
j_{k} \\ \pmb{c}_{b} \\
\end{bmatrix}-\mathfrak{C}_{j_{k}b}\begin{bmatrix}
j_{k} \\ \pmb{c}_{a} \\
\end{bmatrix}\right)+\sum_{\substack{s=1 \\ s\neq k}}^{r}\mathfrak{C}_{j_{k}j_{s}}\det\begin{bmatrix}
j_{k} & j_{s} \\
\pmb{c}_{a} & \pmb{c}_{b} \\
\end{bmatrix}=0\,,
\end{align}
such that the first sum in \eqref{eq:id2} becomes:
\begin{align}
&\quad\sum_{k=1}^{r}\frac{[aq][j_{k}b]}{[j_{k}q]}\left(\mathfrak{C}_{aj_{k}}\det\begin{bmatrix}
j_{k} \\ \pmb{c}_{b} \\
\end{bmatrix}-\mathfrak{C}_{j_{k}b}\begin{bmatrix}
j_{k} \\ \pmb{c}_{a} \\
\end{bmatrix}\right)\nonumber\\
&=-\sum_{k=1}^{r}\sum_{\substack{s=1 \\ s\neq k}}^{r}\frac{[aq][j_{k}b]}{[j_{k}q]}\mathfrak{C}_{j_{k}j_{s}}\det\begin{bmatrix}
j_{k} & j_{s} \\
\pmb{c}_{a} & \pmb{c}_{b} \\
\end{bmatrix}\nonumber\\
&=-\frac{1}{2}\sum_{k=1}^{r}\sum_{\substack{s=1 \\ s\neq k}}^{r}\left(\frac{[aq][j_{k}b]}{[j_{k}q]}-\frac{[aq][j_{s}b]}{[j_{s}q]}\right)\mathfrak{C}_{j_{k}j_{s}}\det\begin{bmatrix}
j_{k} & j_{s} \\
\pmb{c}_{a} & \pmb{c}_{b} \\
\end{bmatrix}\nonumber\\
&=-\frac{1}{2}\sum_{k,s=1}^{r}\frac{[j_{k}j_{s}][aq][bq]}{[j_{k}q][j_{s}q]}\mathfrak{C}_{j_{k}j_{s}}\det\begin{bmatrix}
j_{k} & j_{s} \\
\pmb{c}_{a} & \pmb{c}_{b} \\
\end{bmatrix}\,,
\end{align}
which exactly cancels the second term of \eqref{eq:id2}. Now we have completed the proof of $\mathcal{A}_{ab}=0$, and thus \eqref{eq:Aab}.

\bibliographystyle{JHEP}
\bibliography{Refs}

\end{document}